\documentclass[%
 reprint,
superscriptaddress,
 amsmath,amssymb,
 aps,
pra,
floatfix,
]{revtex4-2}
\usepackage{textcomp}
\usepackage{graphicx}
\usepackage{dcolumn}
\usepackage{bm}
\usepackage[colorlinks=true,linkcolor=blue,urlcolor=blue,citecolor=blue]{hyperref}
\usepackage{comment}

\usepackage[dvipsnames]{xcolor}
\begin{document}

\preprint{APS/123-QED}

\title{ Decoding \textit{active} force fluctuations from spatial trajectories of active systems}

\author{Anisha Majhi}
\affiliation{Department of Physical Sciences, Indian Institute of Science Education and Research, Kolkata 741246, India} 
\author{Biswajit Das}
\email{bd18ip005@iiserkol.ac.in}
\affiliation{Department of Physical Sciences, Indian Institute of Science Education and Research, Kolkata 741246, India}
\author{Subhadeep Gupta}
\thanks{Present Address: London School of Hygiene and Tropical Medicine, Keppel Street, London WC1E 7HT, UK}
\affiliation{Department of Biological Sciences, Indian Institute of Science Education and Research, Kolkata 741246, India}
\author{Anand Dev Ranjan}
\affiliation{Department of Physical Sciences, Indian Institute of Science Education and Research, Kolkata 741246, India}
\author{Amirul Islam Mallick}
\affiliation{Department of Biological Sciences, Indian Institute of Science Education and Research, Kolkata 741246, India}
\author{Shuvojit Paul}
\email{shuvojit.paul@krc.edu.in}
\affiliation{Department of Physics, Kandi Raj College, Kandi, Murshidabad, West Bengal, 742137, India}
\author{Ayan Banerjee}
\email{ayan@iiserkol.ac.in}
\affiliation{Department of Physical Sciences, Indian Institute of Science Education and Research, Kolkata 741246, India}


\date{\today}

\begin{abstract}

Mesoscopic active systems exhibit various unique behaviours - absent in passive systems - due to the forces generated by the corresponding constituents by converting their available free energies. However, estimating these forces - which are also stochastic and remain intertwined with the thermal noise - is especially non-trivial. Here, we introduce a technique to extract such fluctuating active forces acting on a passive particle immersed in an active bath with high statistical accuracy by filtering out the related thermal noise. We first test the efficacy of our method under numerical scenarios with different types of activity, and then apply it to the experimental trajectories of a microscopic particle (optically) trapped inside an active bath consisting of motile \textit{E.Coli.} bacteria. We believe that our simple yet powerful approach, which appears agnostic to the nature of the active force,  should enable accurate measurement of force dynamics in living matter and potentially allow direct but reliable estimation of key thermodynamic parameters such as heat, work, and entropy production.


\end{abstract}
\maketitle


\section{\label{sec:level1}Introduction:}
Microscopic active systems, such as motile bacteria, active Brownian particles, etc., can produce their own forces and motion by exploiting their free energies (originating from internal physio-chemical processes) or utilizing proper external perturbations ~\cite{fodor2016far,ramaswamy2017active,vrugt2024review}. Due to these self-generated forces, these systems inherently stay away from equilibrium and can exhibit several complex collective phenomena~\cite{bechinger2016active,bowick2022symmetry}. The behavior of embedded passive particles in fluids containing these active systems, i.e., \textit{active fluids}, has also grown in interest enormously in recent times~\cite{wu2000particle,leptos2009dynamics,argun2016non,ortileb2019statistics,maggi2014generalized,dor2022passive,reichert2021tracer,kushwaha2023phase,singh2024anomalous,caprini2024emergent}. In such fluids, passive particles interact differently than in \textit{passive fluids} and show several intriguing effects, including non-gaussian and enhanced diffusion~\cite{leptos2009dynamics,ortileb2019statistics,lagarde2020colloidal,nordanger2022anisotropic,santra2023dynamical,goswami2024anomalous}, occurrences of unidirectional motion of asymmetric passive particles~\cite{di2010bacterial, sokolov2010swimming}, and the emergence of long-range effective interactions between particles~\cite{zaeifi2017effective,baek2018generic}. The microscopic reason behind these behaviors is clearly the force acting on these passive particles from the active constituents in the fluid (this force can be called ``\textit{active force}"). The time-averaged nature of \textit{active force} has been studied fairly elaborately, both theoretically and experimentally ~\cite{ortiz2005transport,angelani2011effective,paul2022force,jayaram2023effective,shea2024force}, and has displayed various interesting properties. For example, the average active force between passive particles shows wall curvature dependency~\cite{solon2015pressure,smallenburg2015swim}, Casimir-like effects~\cite{ray2014casimir}, tunability depending on active constituent density~\cite{ni2015tunable}, oscillating nature with distance ~\cite{paul2022force} and constraint dependency ~\cite{liu2020constraint}. However, only a few works exist to comprehend the fluctuation properties of the \textit{active force}. Indeed, to explain the enhanced positional fluctuations and multiple time scales (different from the diffusive time scaling $\sim t$) present in the mean squared displacement function (MSD) of passive particles in \textit{active fluid}s, often the statistical nature of the underlying \textit{active force} is considered as Gaussian distributed and exponentially correlated ~\cite{maggi2014generalized}. Nevertheless, such an assumption fails to explain the non-Gaussian properties repeatedly found in the positional fluctuations of passive particles immersed in active fluids ~\cite{leptos2009dynamics,argun2016non, ortileb2019statistics}. It has been recently shown that an \textit{active force} model with \textit{poissonian active kicks} can be useful in this regard~\cite{di2024brownian}. Clearly, the knowledge of the statistical nature of \textit{active force} is extremely limited to date.

Further, without complete knowledge of \textit{active force} fluctuations, the measurement of the thermodynamical parameters related to systems of passive particles in active fluids is rather insubstantial. It is known that, without the trajectory of force fluctuations, the irreversibility of the positional trajectories of passive particles measured using KL-divergence can be used as the only parameter to understand the thermodynamic effects of such systems~\cite{roldan2021quantifying}. However, KL-divergence generally provides an extreme lower bound of the total entropy production rate of the system. The availability of active force fluctuations could certainly provide the correct estimation of the total entropy production rate~\cite{dabelow2019irreversibility,manikandan2021quantitative,das2024irreversibility}. Moreover, thermodynamical quantities such as work and heat can also then be estimated using the usual trajectory energetics ~\cite{sekimoto1998langevin,das2023enhanced}. Additionally, a full time series of \textit{active force} could offer an unfiltered view of the energy exchanges occurring at every moment, providing granular insight into the system that not only deepens our understanding of non-equilibrium processes but also bridges the gap between experimental observations and theoretical models in active matter physics. Also, in living systems, where active processes are central to basic functioning, a method providing an avenue for directly probing the forces driving cellular motion and interactions could have far-reaching implications for understanding processes such as membrane tension heterogeneity~\cite{Ghosh2025.02.17.638627}, cell migration~\cite{chen2023understanding, saraswathibhatla2020spatiotemporal}, microbial dynamics~\cite{sabass2017force}, and tissue development~\cite{heisenberg2013forces}.
However, to the best of our knowledge, no work is present till date that measures the whole fluctuating \textit{active force} trajectory and the corresponding statistical properties directly, when such a force acts on a passive particle in active fluids. The reason behind this is understandable -- the thermally originated random force also remains added to the random \textit{active force}, and one needs to correctly filter the former out. Here, we propose a technique to infer the active force fluctuations from the positional fluctuations of a passive probe confined in an active environment. We demonstrate the efficacy of the model by both simulations and experiments, and observe that the model yields very good results - especially when the active force component is comparable or higher compared to the thermal force.

\begin{figure}[h]
    \includegraphics[width=0.5\textwidth]{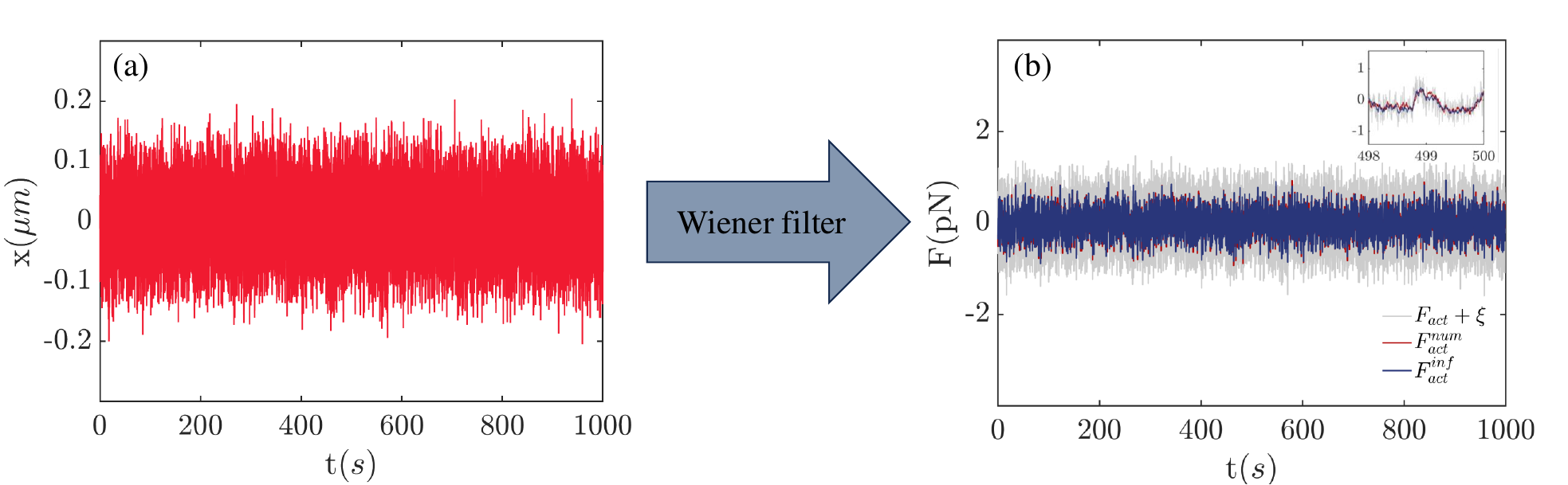}
    \caption{Numerically generated position fluctuations of an optically trapped passive spherical particle are fed to the filter for obtaining filtered active force trajectory. (a) Positional trajectories of the particle. (b) Inferred active force ($F_{act}^{inf}$, in color blue) along with total force ($F_{act}^{num} +\xi$, in gray) and the active force used for the numerical simulation ($F_{act}^{num}$, in red) with time. Inset: expanded force trajectories to understand the efficacy of the method.}
    \label{Filt_traj}
\end{figure}
 We choose a model system that is easily realizable in experiments to demonstrate our method; that is a harmonically trapped microscopic particle immersed in an active bath (see the Fig.~\ref{Experiment}(a)). The underlying dynamics of the particle localized inside the bath with an external harmonic potential of stiffness $k$ can be described by the \textit{Langevin} equation, 
\begin{equation}
    \gamma\dot{x}(t) + kx(t) = F(t) = \xi(t) + F_{act}(t),
    \label{eq1}
\end{equation}
\begin{figure*}[t]
    \centering
    \includegraphics[width=1\textwidth]{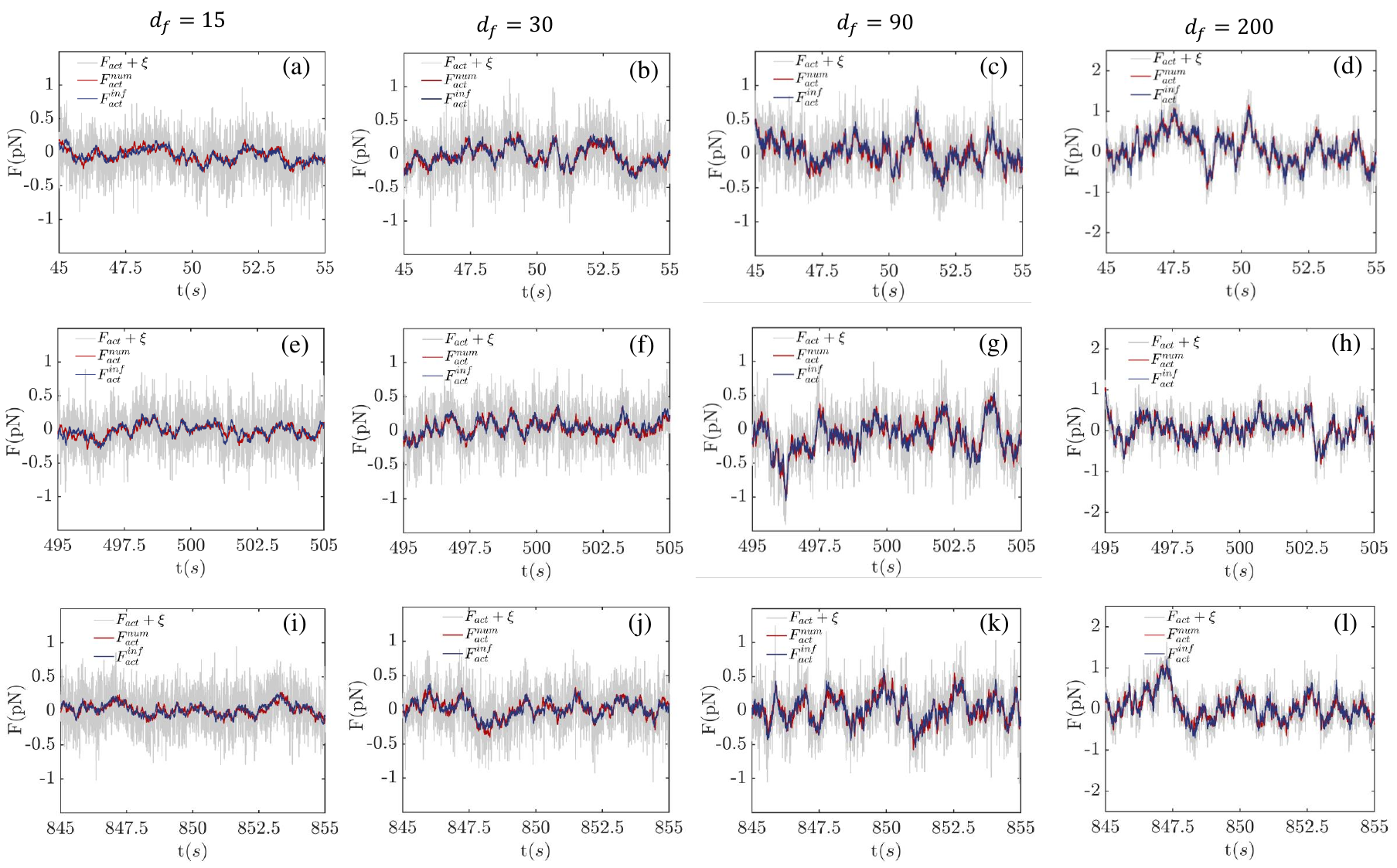}
    \caption{Magnified force trajectories from the numerical scenario I for different $d_f$ values to observe real-time correspondence between the (input) numerical active force and its inferred output. The time intervals are randomly chosen from the beginning, middle and end of the total sampling time. Left to right:  The columns represent the enlarged force trajectories for $d_f$ values 15, 30, 90 and 200, respectively at different time periods.}
    \label{realtime}
\end{figure*}
where, $x(t)$ denotes the one-dimensional position of the particle at time $t$, and $\gamma ~(=6\pi\eta a)$ represents the related friction coefficient which depends on the radius of the microparticle ($a$) and the viscosity of the medium ($\eta$). The net force acting on the particle from the bath $F(t)$ is the sum of the thermally originated delta correlated random force $\xi(t)$, i.e. $\langle\xi(t)\rangle = 0$, $\langle\xi(t)\xi(t^\prime)\rangle = 2k_B T\gamma \delta(t-t^\prime)$ and the stochastic force originating from the active constituents (e.g., motile bacteria, active Brownian particles, etc.) present only inside the bath, i.e., $F_{act}(t)$. Here $k_B$ denotes \textit{Boltzmann} constant and $T$ is the temperature of the bath. Due to the linearity of the \textit{Langevin} equation, $F(t)$ can be conveniently calculated from the position $x(t)$ of the trapped particle if $k$ and $\gamma$ are known (see Eq.~\ref{eq1}). Here, we show that $\xi(t)$ from the net force $F(t)$ can be filtered out to infer the active force $F_{act}(t)$ exclusively. We first test our technique using numerical scenarios and then apply it to the experimentally recorded trajectories of a microscopic particle optically trapped inside motile bacteria baths of different concentrations.\\

The article is organized as follows: We first describe our technique in Section~\ref{sec:level2}, after which we evaluate its efficacy with a numerical scenario in Section~\ref{sec:level3}. Subsequently, in Section~\ref{sec:level4}, we employ this technique to infer and comprehend active forces acting on a passive particle optically trapped inside an active bath comprised of motile \textit{E.Coli.} cells.  Lastly, we conclude our article with discussions and future directions in Section~\ref{sec:level5}. 
\begin{figure*}[t]
    \centering
    \includegraphics[width=1\textwidth]{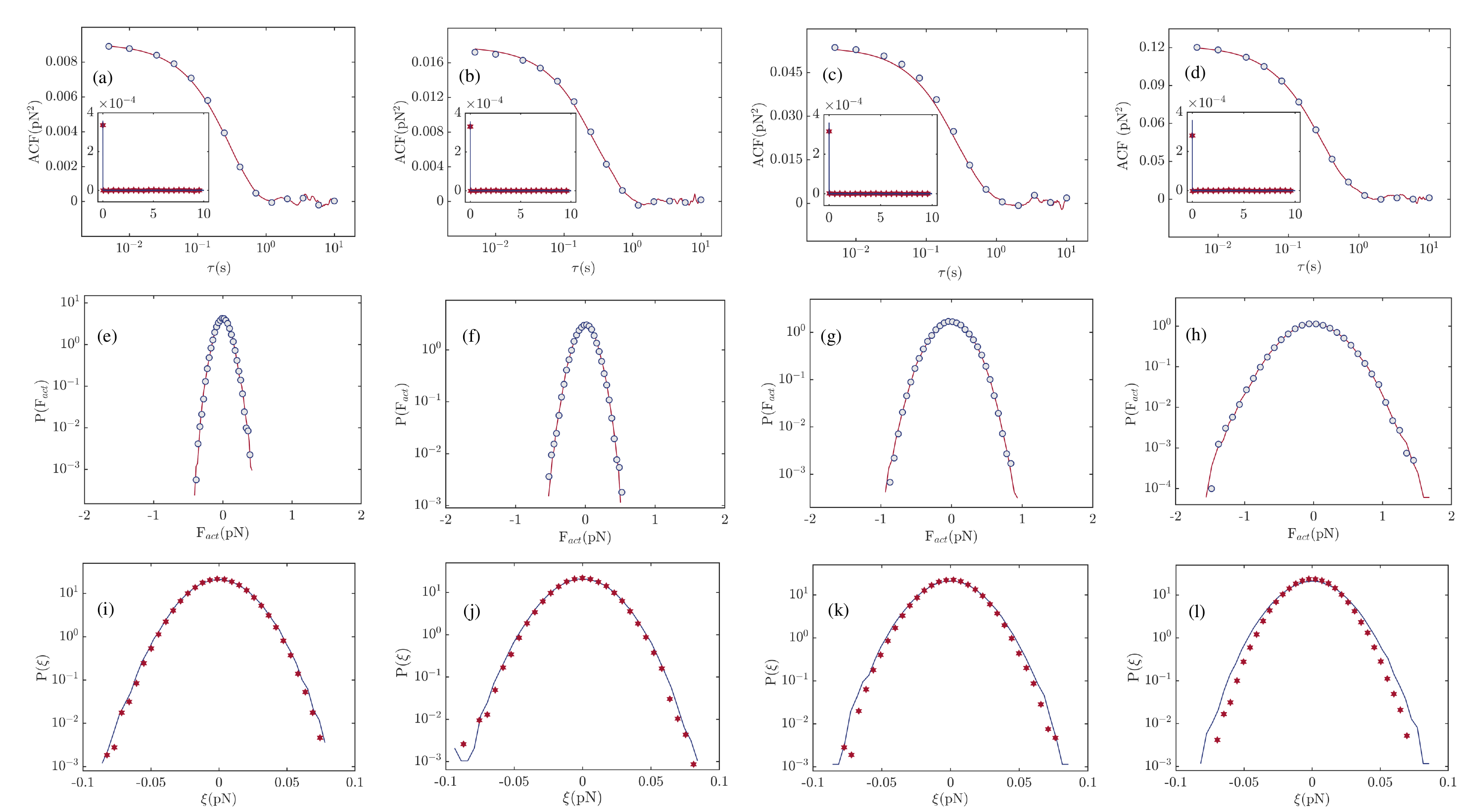}
    \caption{ \textbf{Top panel:} Autocorrelation function (ACF) plotted for numerically generated [\textit{input}] active force (line) and inferred active force (circle) by filtering. Inset: ACF of the thermal force of numerical origin plotted against corresponding inferred (star) result. (a)-(d) are in the order of increasing activity strength, where $d_f= 15, 30, 90, 200$ respectively. Corresponding normalised errors ($\frac{\text{MSE}}{Var(F_{act})}$) are computed as, $\sim 0.35, 0.26, 0.17, 0.11$  for $d_f=15, 30, 90, 200$ respectively. \textbf{Middle panel:} (e)-(h) shows the probability distribution function (PDF) of active forces (numerically generated (line) and numerically inferred (circle)) corresponding to increasing values of $d_f$ as mentioned. \textbf{Bottom panel:} (i)-(l) PDF of the thermal force of numerical origin plotted with corresponding inferred (star) result in ascending order of activity strength. [Parameters that are fixed in the numerical simulation: $k_B = 1.38\times10^{-23}~$J/K, $T = 310~$K, $a=2.5~\text{\textmu m}$, $k=7~\text{pN/\textmu m}$, $\eta=8.9\times10^{-4}~$Pa.s, $\tau_e = 0.3~$s]}
    \label{fig:fig2}
\end{figure*}
\section{\label{sec:level2}Method: Based on Wiener Filter} Assuming the process to be stationary, we use the time domain Wiener filter to estimate the active part of the fluctuating time series of the force, i.e., $F_{act}(t)$ acting on the trapped probe particle ~\cite{wiener1949extrapolation}. Note that Brownian motion itself is a Wiener process \cite{wiener1923differential}, and the Wiener filter has been extensively used to remove random noise from a variety of data in image processing, signal processing, control systems, and digital communications, where the noise is additive \cite{boulfelfel1994three}.  In our case, as the properties of the thermally originated component of the force $\xi(t)$ are known, with that also being uncorrelated to the active component $F_{act}(t)$, the Wiener filter can be efficiently employed. As described in the Eq.~\ref{eq1}, the resultant force on the probe can be written as the following:
\begin{align}
    F(t)=F_{act}(t)+\xi(t),
    \label{eq2}
\end{align}
and, for this system, $F(t)$ is related to the trajectory of the probe $x(t)$ as
\begin{align}
    F(t)=\gamma \dot{x}(t)+kx(t).
    \label{eq3}
\end{align}
For discrete measurements taken at \(N\) time points, we denote the recorded net force as the vector $\mathbf{F} = \left[ F_1, F_2, \dots, F_N \right]$.
To extract the active force component from \(\mathbf{F}\), we employ the Wiener filter. This filter is a linear minimum mean-square error (LMMSE) estimator that constructs an estimate of the active force in the form
\begin{equation}
    \hat{\mathbf{F}}_{act} = \mathbf{w}^T \mathbf{F},
\end{equation}
where \(\mathbf{w}\) is a vector of weights and the superscript \(T\) denotes the transpose. The mean-square error (MSE) between the true active force \(\mathbf{F}_{act}\) (expressed as a vector) and its estimate is defined as
\begin{align}
    \text{MSE} = \left\langle \left( \mathbf{F}_{act} - \mathbf{w}^T \mathbf{F} \right)^2 \right\rangle,
    \label{eqMSE}
\end{align}
where \(\langle \cdot \rangle\) denotes a average over time. \\
To obtain the optimal weight vector \(\mathbf{w}\) (denoted as \(\mathbf{w}^*\)) that minimizes the MSE, we set the gradient with respect to \(\mathbf{w}\) to zero:
\begin{equation}
    \nabla_{\mathbf{w}}\,\text{MSE} = \nabla_{\mathbf{w}} \left\langle \left( \mathbf{F}_{act} - \mathbf{w}^T \mathbf{F} \right)^2 \right\rangle = \mathbf{0}.
\end{equation}
This condition yields
\begin{equation}
    -\left\langle \mathbf{F}_{act}\, \mathbf{F} \right\rangle + \left\langle \mathbf{F}\, \mathbf{F}^T \right\rangle\, \mathbf{w} = \mathbf{0}.
\end{equation}
We now define the \textit{cross-correlation} vector between the active force and the net force as
\begin{equation}
    \mathbf{r} = \left\langle \mathbf{F}_{act}\, \mathbf{F} \right\rangle \nonumber,
\end{equation}
and the \textit{auto-correlation} matrix of the net force as
\begin{equation}
    \mathbf{R} = \left\langle \mathbf{F}\, \mathbf{F}^T \right\rangle \nonumber.
\end{equation}
Thus, the optimal weight vector is given by
\begin{equation}
    \mathbf{w}^* = \mathbf{R}^{-1} \mathbf{r} \nonumber,
\end{equation}
and the Wiener filter estimate of the active force becomes
\begin{equation}
    \hat{\mathbf{F}}_{act} = \mathbf{w}^{*T} \mathbf{F}.
\end{equation}
An important aspect of this method is that the thermal noise \(\xi(t)\) is statistically independent of the active force \(F_{act}(t)\), i.e.,
\begin{equation}
    \left\langle \mathbf{F}_{act}\, \xi \right\rangle = \mathbf{0} \nonumber.
\end{equation}
Consequently, the cross-correlation vector simplifies to the auto-correlation of the active force:
\begin{equation}
    \mathbf{r} = \left\langle \mathbf{F}_{act}\, \mathbf{F}_{act} \right\rangle \nonumber.
\end{equation}
Since the auto-correlation of the net force is the sum of the auto-correlations of the active force and the thermal noise,
\begin{equation}
    \left\langle \mathbf{F}\, \mathbf{F} \right\rangle = \left\langle \mathbf{F}_{act}\, \mathbf{F}_{act} \right\rangle + \left\langle \xi\, \xi \right\rangle \nonumber,
\end{equation}
we can also write
\begin{equation}
    \mathbf{r} = \left\langle \mathbf{F}\, \mathbf{F} \right\rangle - \left\langle \xi\, \xi \right\rangle.
\end{equation}
The thermal noise \(\xi(t)\) is well characterized by temperature \(T\) of the system. Specifically, for a system in thermal equilibrium, the auto-correlation of the thermal noise is given by
$\left\langle \xi\, \xi \right\rangle = \left[ 2 k_{B} T\, \gamma,\, 0,\, 0,\, \dots,\, 0 \right]$
where \(k_{B}\) denotes the Boltzmann constant.
Further, as $\textbf{F}$ can be calculated easily from the trajectory of the probe, $\textbf{R}$ and $\textbf{r}$ can also be determined.
It is also because of this reason that it is advantageous to use the Wiener filter in this application, as by construction, it separates the forces based on statistical properties, unlike other non-adaptive methods (e.g. low-pass filters, moving average filters) which often fail to distinguish different stochastic forces in overlapping frequency ranges~\cite{diniz1997adaptive}.

\section{\label{sec:level3}Numerical Scenario I:  Ornstein Uhlenbeck noise as active force}
First, we test our technique with numerical trajectories of a colloidal particle optically trapped in an active bath. The active force present in the bath is modelled as an exponentially correlated Ornstein-Uhlenbeck (OU) stochastic force with the correlation of the form: $\langle F_{act}(t) F_{act}(t^\prime)\rangle = \gamma^2 D_e/\tau_e \exp(-(t-t^\prime)/\tau_e)$, as it is often used for this purpose~\cite{maggi2014generalized, dabelow2019irreversibility}. To implement it numerically, the active force ($F_{act}(t) \equiv \gamma \xi_{act}(t)$) can be represented by a stochastic variable ($\xi_{act}(t)$) following a linear \textit{Langevin} equation with drift $-1/\tau_e$ and noise coefficient $\sqrt{2D_e}/\tau_e$~\cite{dabelow2019irreversibility,das2023enhanced,das2024irreversibility}. The strength of the active force then can be denoted as $D_e$, which can be termed as the \textit{active diffusion coefficient}. The correlation time of the force is denoted by $\tau_e$.  We introduce a parameter $d_{f}$ as the ratio of the active ($D_{e}$) and thermal ($D_{0} = k_B T/\gamma$) diffusion coefficients, i.e., $d_{f}=D_{e}/D_{0}$. Clearly, a higher value of $d_f$ indicates a stronger contribution of the active force. 
\begin{figure*}
    \centering
    \includegraphics[width=0.9\textwidth]{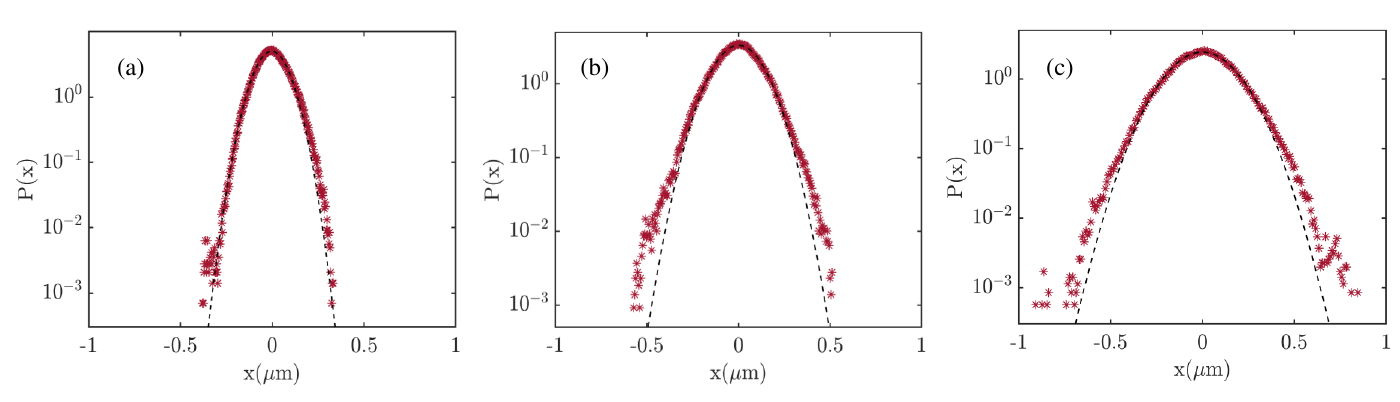}
    \caption{(a)-(c) are the positional probability density functions corresponding to the numerical trajectories of a passive probe in a bath of ABPs with increasing average propulsion speeds having values 50$~\text {\textmu m/s}$, 100$~\text {\textmu m/s}$, and 200$~\text {\textmu m/s}$ respectively. The deviation of PDF from the Gaussian profile (dotted line) highlights the non-Gaussian nature of the underlying fluctuations. Stronger activity conditions lead to more pronounced breakdown of Gaussian statistics in the system.}
    \label{ABP_pdf_x}
\end{figure*}
We vary the value of $d_{f}$, keeping all other parameters fixed, and generate trajectories of the trapped particle, i.e., $x(t)$ for time 1000 secs with numerical time step $0.005~$s.  To generate these trajectories, the \textit{Langevin} equations mentioned in Eq.(\ref{eq1}) and that related to the active force are used to create a vector autoregressive model~\cite{fricks2009time,paul2021bayesian}  and then simulated using \textit{varm} toolbox of MATLAB~\cite{mathworks_varm}. To match with experiments, we keep the trap stiffness fixed at $k=7~\text{pN/\textmu m}$ and $\eta=8.9\times10^{-4}$ Pa.s.  The net force (active force plus the thermal force) on the particle is calculated following the Eq.~\ref{eq3}. Further, the method described in the section \ref{sec:level2} is followed to filter the thermal part from the net force, and to infer the active component only. Additionally, in the \textit{Appendix} \ref{filter_optimization}, we have shown that a priori knowledge of the system time scale and thermal noise strength are not necessary for this method, which is an important highlight of this process. Indeed, both can be changed in the presence of many interacting particles in the bath. An initial presumption can be iteratively converged to the right values of these parameters with fairly high accuracy. One of the filtered trajectories is shown in Fig.~\ref{Filt_traj}(b), in which we describe our method pictorially. Note that the thermal force can be conveniently calculated by subtracting the inferred active force from the net force, and the corresponding statistical natures can also be calculated. The \textit{auto-correlation function} (ACF) and the \textit{probability density function} (PDF) of the estimated active force and the thermal force along with that of the known exact trajectories are shown in Fig.~\ref{fig:fig2} for four values of $d_{f}$. The related errors of estimations are performed from the minimum of the MSE (Eq.~\ref{eqMSE}), and their values normalized by the variance of the simulated active force are shown in Fig.~\ref{fig:fig2}. As expected, the accuracy of the filter monotonically increases with the increase of $d_{f}$. However, it is primarily limited by the finite length of the trajectory, which is further discussed in the \textit{Appendix}~\ref{filter_accuracy}. Importantly, we find that the statistical parameters match considerably well for all the $d_{f}$ values, as shown in Fig.~\ref{fig:fig2}. Note that the delta-correlation function and the Gaussian probability distribution function of the noise should be independent of $d_{f}$, and thus, they too should be utilized to check the correctness of the estimation (\textit{see Appendix} \ref{filter_optimization}).

\begin{figure*}
    \centering
    \includegraphics[width=0.9\textwidth]{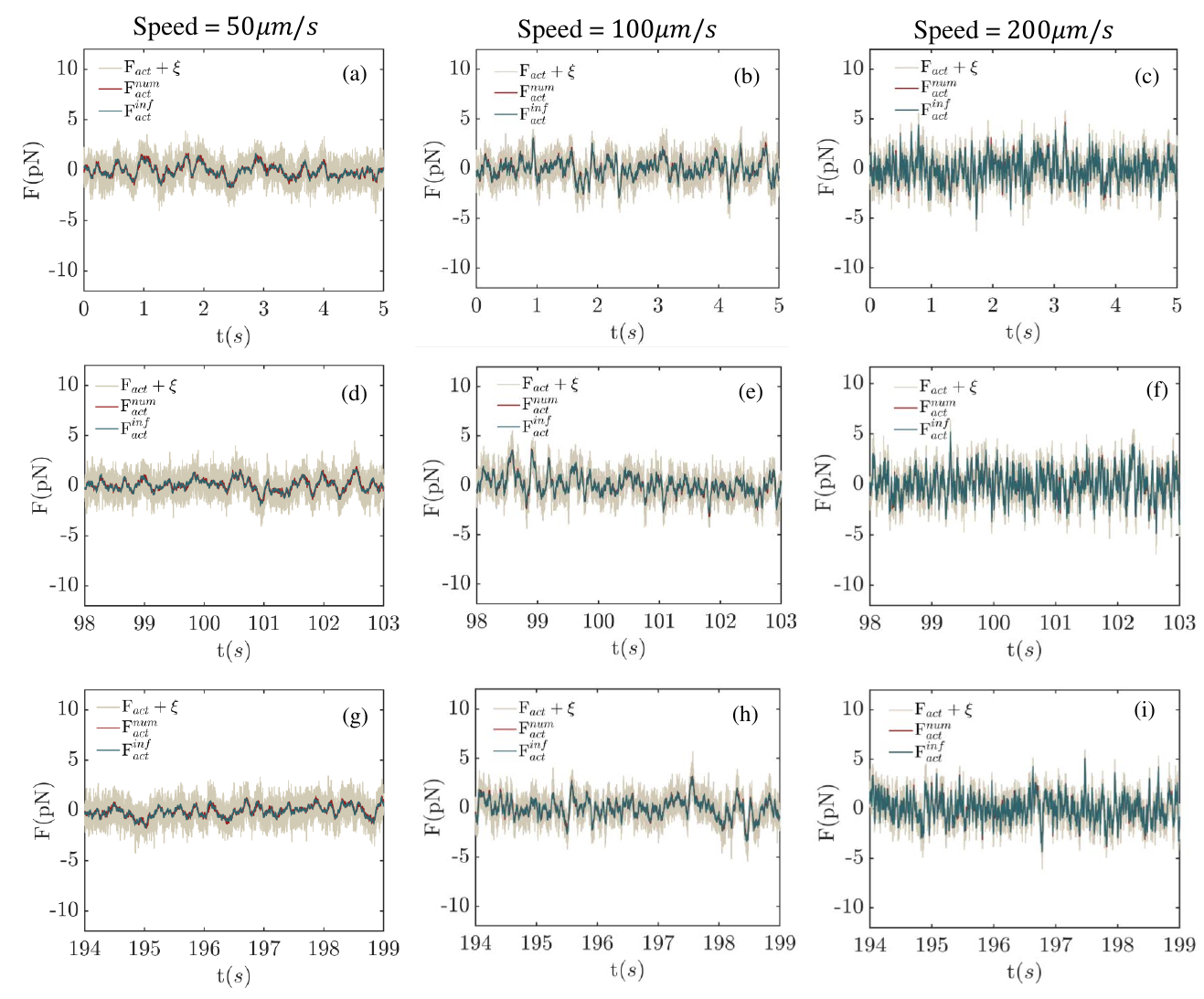}
    \caption{The trajectories of the constituent forces acting on a passive probe immersed in a bath of ABPs are simulated. To assess the accuracy of the active force inference, real-time comparison between the numerical input active force ($F_{act}^{num}$) and inferred active force ($F_{act}^{inf}$) are observed within randomly selected time windows. $\xi$ represents the thermal force prevalent in the system. Each column corresponds to the numerical simulations with different average propulsion speeds of the ABPs.}
    \label{ABP_realtime}
\end{figure*}

\begin{figure*}
    \centering
    \includegraphics[width=0.9\textwidth]{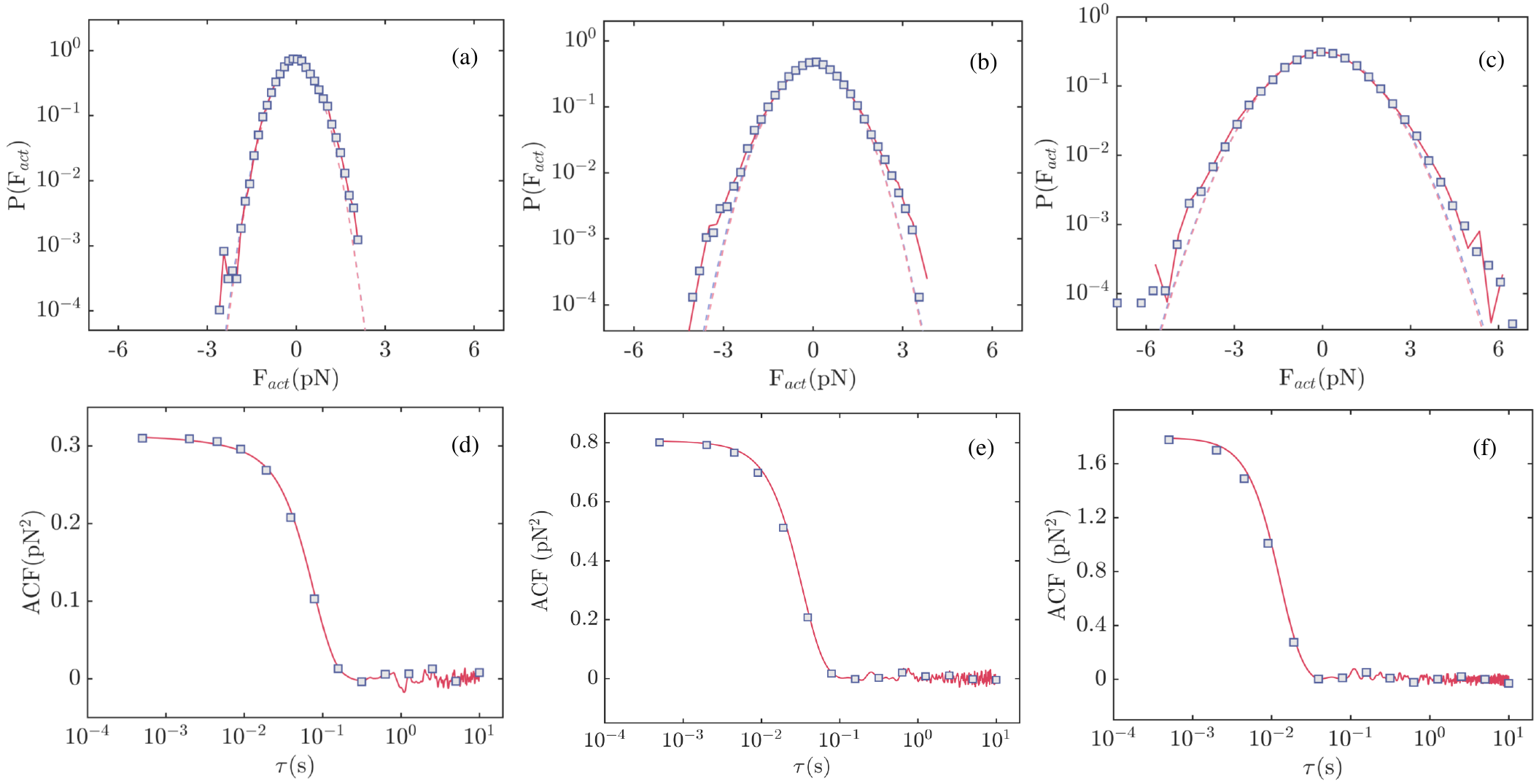}
    \caption{Statistical properties of the numerically generated active forces (line) are plotted against their inferred estimates (square) for comparison.  (a)–(c): Positional probability density functions (PDFs) of the associated active forces. The extent to which the force PDFs deviate from a Gaussian profile underscores the non-Gaussian nature of the intrinsic force fluctuations. This deviation becomes more pronounced at higher activity strengths. (d)–(f): Autocorrelation functions (ACFs) of the active forces experienced by a passive probe in a bath of ABPs with increasing average propulsion speeds of 50$~\text {\textmu m/s}$, 100$~\text {\textmu m/s}$, and 200$~\text {\textmu m/s}$, respectively.} 
    \label{ABP_force_pdf_acf}
\end{figure*}

\begin{figure*}
    \includegraphics[width=1\textwidth]{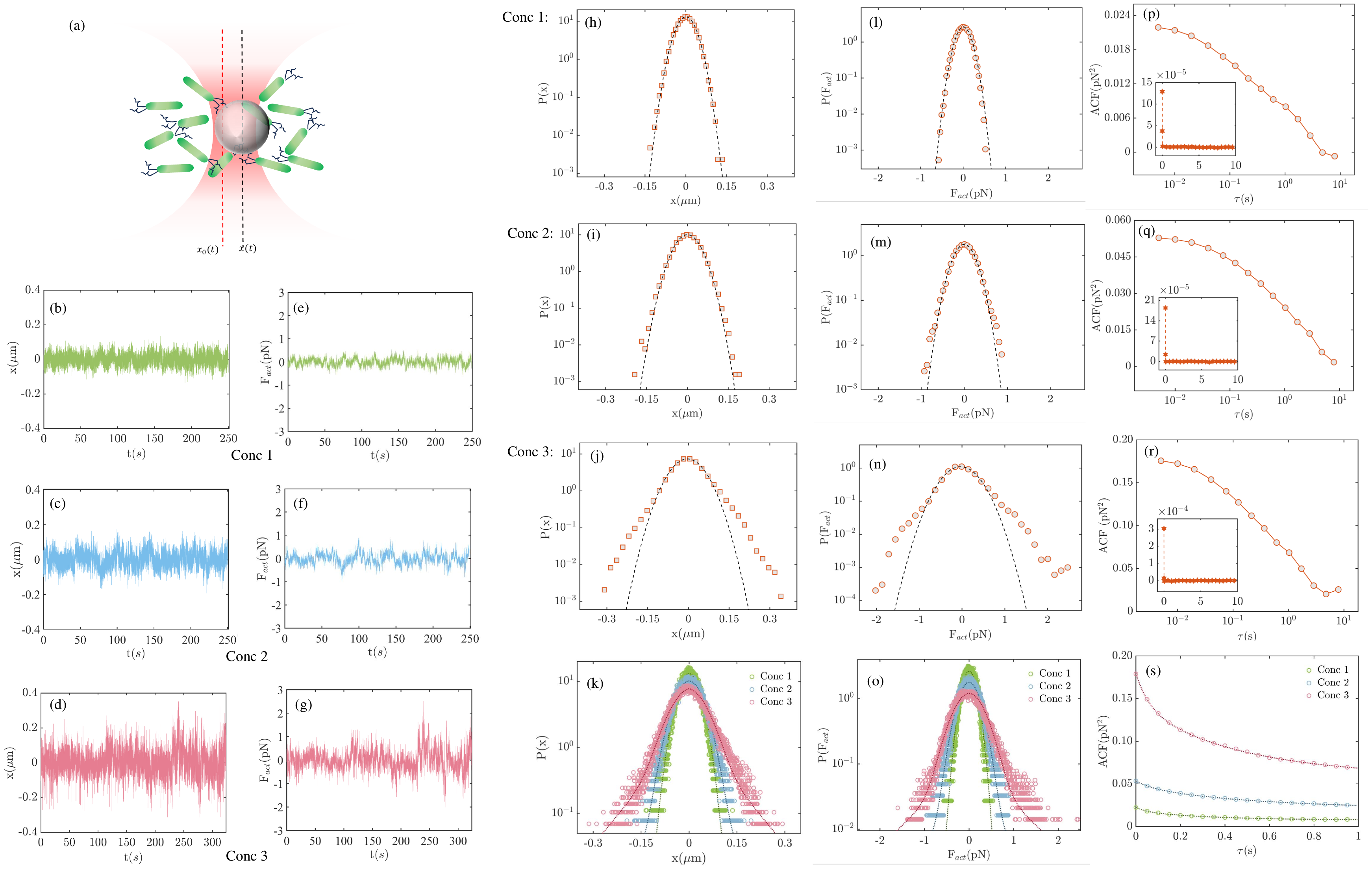}
    \caption{(a) Schematic of an optically confined passive probe in motile \textit{E.coli} bath. Experiments are performed for three different concentrations of active cells. Conc 1: $2\times10^5$ CFU, conc 2: $1\times10^6$ CFU, conc 3: $5\times10^6$ CFU (where CFU stands for colony forming units). (b)-(d) Positional trajectories of the passive probe immersed in a bath of active \textit{E.coli} cells at varying cell concentrations. (e)-(g) Trajectories of the effective active forces inferred from the positional trajectories for the respective bath concentrations. (h)-(j) The probability distribution functions ($P(x)$) corresponding to the experimentally recorded position fluctuations of the passive probe in the active \textit{E.coli} bath with different concentrations are shown. The dotted lines indicate the Gaussian core of $P(x)$.  (k) The deviation from Gaussian characteristics of $P(x)$ at higher bacterial concentrations is indicated through the empirical fits. Corresponding function is described as $P(y)$ in the main text. (i)-(n)  The probability distribution functions of the inferred active forces ($P(F)$) from the experimental trajectories for different bacterial concentrations are plotted. (o) $P(F)$ corresponding to all the concentrations are empirically fitted with non-Gaussian functions, described as $P(y)$ in the main text. (p)-(r) Autocorrelation function (ACF) of the inferred active forces are shown and ACFs of the thermal forces extracted from corresponding trajectories are shown in insets. (s) ACFs of the inferred active forces are fitted with double-exponential functions with different decay rates.} 
    \label{Experiment}
\end{figure*}

\section{\label{sec:ABP bath} Numerical Scenario II:  Passive particle in an ABP bath}
Next, we consider a system with a colloidal probe harmonically confined in a bath of active Brownian particles (ABPs)~\cite{shea2024force}. The system is realised numerically by simulating the $2D-$ Langevin dynamics of $N=100$ identical ABPs with average self-propulsion velocity $v_0$ in a thermal bath of temperature $T=310\ K$. The passive particle -- five times larger in size than the ABPs -- is confined in a harmonic potential of stiffness $k = 7\ pN/\mu m$ and follows separate \textit{Langevin} dynamics in the presence of both active particles and thermal fluctuations. The details of the numerical simulations with the governing dynamical equations are provided in \textit{Appendix}~\ref{ap:simulation}. At the coarse-grained level, the kinetics of the passive particle is a resultant motion due to the simultaneous perturbation of the active and thermal fluctuations, and the spatio-temporal trajectories of the passive particle should contain that information. Our goal is to test the effectiveness of the proposed technique by extracting  the total active force fluctuations from simulated trajectories of the passive one.

We proceed by varying the self-propulsion speed of the ABPs as 50$~\text {\textmu m/s}$, 100$~\text {\textmu m/s}$, and 200$~\text {\textmu m/s}$ to mimic the different levels of activity. 
In particular, we observe that the position fluctuations of the passive particle deviate from Gaussian characteristics with this concentration of the ABPs, especially at higher propulsion speeds of APBs (Fig.~\ref{ABP_pdf_x}(a)-(c)). This is in contrast with the previous example of considering OU noise as the response of such an active bath that always results in Gaussian position fluctuations of the passive probe. Nevertheless, we apply the same technique on the trajectories of the passive probe and filter out the thermal fluctuations to learn about the total active fluctuations. We show that the filtered active force fluctuations match reasonably well with the effective fluctuating active force generated due to the interaction of the ABPs with the passive particle at different levels of activity - characterized through the average propulsion speed of ABPs (Fig.~\ref{ABP_realtime}). Moreover, the statistical properties of the extracted active force, such as the probability distribution with prominent non-Gaussian features (Fig.~\ref{ABP_force_pdf_acf}(a)-(c)) and the exponential nature of the autocorrelation functions (Fig.~\ref{ABP_force_pdf_acf}(d)-(f)), are indeed prevalent and are in agreement with the nature of the input active force. \\

After establishing the fidelity of our proposed technique with two separate numerical scenarios, we next proceed to employ this to an Optical Tweezers-based experimental system.
\section{\label{sec:level4}Experiment: Passive particle in an active bacterial bath}

We design an Optical Tweezers-based experimental arrangement to test the efficacy of our method and to infer the stochastic force from the experimental position fluctuations of a trapped microparticle. To create the optical trapping potential, we use the modular Optical Tweezers system (OTM, ThorLab) in which a Gaussian laser beam of wavelength $\lambda = 1064$ nm is tightly focused by an objective lens of high numerical aperture (NA = 1.4, 100X, oil-immersion) installed in a conventional inverted microscope (NIKON). Silica microparticles (diameter $5 ~\text{\textmu m}$) - sparsely dispersed in a bath of motile \textit{E.coli} cells - are trapped inside a sample holder placed on a temperature-controlled stage. The sample holder is prepared by attaching a Polyethylene glycol (PEG) coated glass coverslip at the bottom of an aluminium plate with a circular hole. The top surface is then sealed by attaching another coverslip, creating a circular cavity of diameter $14~$mm and height $1~$mm. All edges and contact points between the aluminum plate and glass coverslips are ensured to be perfectly sealed to eliminate air contact. PEG coating prevents sticking and accumulation of \textit{E.coli} cells at the bottom of coverslips, which would hamper the uniformity of the bacterial concentration over time. The details of preparing the particle-cell mixture are given in \textit{Appendix}~\ref{ecoli_culture}. Note that the size of the enclosed sample chamber is much larger than \textit{E.coli} cells ($\sim 3 ~\text{\textmu m}$) and the trapped probe particle, so that it can be considered isotropic and homogeneous over the length scale of the system -- which further eliminates the possibility of observing persistent rotation of the probe particle inside the bacterial bath. To detect the position fluctuations of the trapped particle, images of the particle are captured at $200$ Hz using a camera (ThorLabs) attached to the microscope for approximately $300~$s. The centre of the particle in each frame is then located using a standard particle tracking algorithm provided in Ref.~\cite{parthasarathy2012rapid}. Note that in our setup, it is entirely safe to assume constant trap stiffness and accurate particle tracking, as the effective beam waist at the focus ($\sim \frac{0.61\lambda}{1.33\text{NA}} \approx 348$ nm, slightly increased by spherical aberration) is larger than the maximum particle fluctuation ($\sim 300$ nm) and the particle size is comparatively huge.

We record the position fluctuations of the trapped particle in three separate concentrations of the bacterial solution ({Conc 1: $2\times10^5$ CFU, Conc 2: $1\times10^6$ CFU, Conc 3: $5\times10^6$ CFU}, where `CFU' stands for `Colony Forming Units'), kept at temperature $T = 37^\circ$C. The stiffness constant of the trap remains fixed at $k = 7.0\pm0.5 ~\text{pN/\textmu m}$ throughout the experiments. The tracked spatio-temporal fluctuations of the confined probe (Fig.~\ref{Experiment}(b) - \ref{Experiment}(d)), along with the inferred active force trajectories (Fig.~\ref{Experiment}(e) - \ref{Experiment}(g)) across the varying bacterial concentrations, reveal increasingly enhanced variations about the mean with rising activity levels. We observe that the fluctuations of the probe in the bacterial bath with concentration $\sim 10^5$ CFU remain Gaussian (Fig.~\ref{Experiment}(h)) and the corresponding active force fluctuations inferred from the trajectory also maintain Gaussian characteristics (Fig.~\ref{Experiment}(l)). However, the probability distributions (PDF) of the position fluctuations of the probe particle are found to be non-Gaussian for the active baths with higher bacterial concentration as shown in Figs.~\ref{Experiment}(i) and \ref{Experiment}(j). The variance of the Gaussian core is also enhanced with the concentration of the bacteria. Note that non-Gaussian PDFs with prominent tails at the edges indicate that the effective active force on the probe particle from the bacterial baths with concentrations $\geq 10^6$ CFU (within our experimental scenarios) may not be Gaussian but rather possess some non-Gaussian characteristics - which are beyond the scope of usual Ornstein-Uhlenbeck (OU) modelling of active forces~\cite{maggi2014generalized}. Indeed, our analysis reveals that non-Gaussian features are clearly present in the corresponding active forces computed from the experimental trajectories  (Figs.~\ref{Experiment}(m) and \ref{Experiment}(n)). 

To account for the appearance of the exponential tails along with the broadening of the Gaussian core, the probability distribution functions of both positional fluctuations ($x(t)$) and active force fluctuations ($F(t)$) are fitted (Fig.~\ref{Experiment}(k) and Fig.~\ref{Experiment}(o)) with a weighted sum of Gaussian and exponential distributions as~\cite{leptos2009dynamics}: $P(y) = \frac{1-f}{\sqrt{2\pi \sigma_g^2}}\exp(-\frac{y^2}{2\sigma_g^2}) + \frac{f}{2\sigma_e} \exp(-\frac{|y|}{\sigma_e}) \label{acf_fit}$. Here, the parameter $\sigma_g$ denotes the standard deviation of the Gaussian distribution, $\sigma_e$ is the decay length of the exponential distribution and $f (0\leq f< 1)$ indicates the fractional contribution of the enhanced displacement or force.   

Moreover, the computed active forces of the bacterial baths are correlated as depicted in Figs.~\ref{Experiment}(p) - \ref{Experiment}(r). To explore it further, we fit the time-dependent autocorrelation functions (ACF) of the corresponding active forces with a double exponential function: $f(t) = A_1 \exp(-t/\tau_1) + A_2 \exp(-t/\tau_2)$, as has also been done in Ref.~\cite{jayaram2023effective}. We assume that the short-time autocorrelation function is a monotonically decaying function with at least two relaxation timescales ($\tau_1$ and $\tau_2$), and the corresponding coefficients ($A_1$ and $A_2$) should be related to the (diffusion) strength of the active forces~\cite{jayaram2023effective}. 

From the empirical fit shown in Fig.~\ref{Experiment}(s), we find that both $\tau_1$ and $\tau_2$ almost remain constant with the bacterial concentrations in the bath and $\tau_1 \ll \tau_2$. Quantitatively, we estimate the longer timescale to be $\tau_2 = (0.46 \pm 0.11)\text{~s}$ over the concentrations - which is comparable to the typical persistent timescales of the motile \textit{E.Coli} cells~\cite{maggi2014generalized}, and the value of $\tau_1$ is at least one order less than $\tau_2$ - which may appear due to the (steric) interaction between the passive probe and the motile bacteria. Furthermore, it is expected that the coefficient related to the slow timescale ($\tau_2$) i.e. $A_2$ should carry the signature of the strength of active diffusion in different concentrations of bacteria.  Taking into account $A_2 \equiv \gamma^2 D_e/\tau_2$, we estimate the active diffusion constants for the three concentrations as,  $D_e$[Conc 1]$=1.97 ~\text \textmu \text m^2/\text s$,  $D_e$[Conc 2]$=8~\text \textmu \text m^2/\text s$ and $D_e$[Conc 3]$=21.42~\text\textmu \text m^2/\text s$. The estimated active diffusion constants are evidently higher than the thermal diffusion ($k_B T/\gamma\sim 0.1~\text\textmu \text m^2/\text s$). The other coefficient $A_1$ is found to be significantly lower than $A_2$ and the effect of this coefficient in estimating the strength of active diffusion can be neglected as $A_1 \tau_1 \ll A_2 \tau_2$. The details of the fitting parameters and the subsequent estimations of the active diffusion constants are tabulated in TABLE~\ref{tab:acf_fit_param}. 
The detailed statistics of the inferred active forces, as presented in TABLE~\ref{tab:acf_fit_param}, not only serve as a diagnostic tool for quantifying active fluctuations, but also reveal important information about the microscopic mechanisms at play. For example, the active force statistics shed light on bacterial motility, density fluctuations, and potential collective behaviors that can be explored in detail. This, in turn, will help bridge the gap between microscopic activity and macroscopic observables, providing a more comprehensive picture of the dynamic behavior of such types of active systems.

\begin{widetext}
    \begin{table*}
    \centering
    \begin{tabular}{|c|c|c|c|c|}
      \hline
       \rule{0pt}{10pt}  & $A_1(\times10^{-3}$pN$^2$) & $\tau_1$(s) & $A_2(\times10^{-3}$pN$^2$) & $\tau_2$(s)\\ [1.2ex]
      \hline
      Conc 1 & $4.342 \pm 0.236 $ & $0.052\pm 0.004$ & $10.05 \pm 0.1920$ & $0.344 \pm 0.013$ \\[0.8ex]
       \hline
      Conc 2 & $7.019 \pm 0.499$ & $0.079 \pm 0.007$ & $25.093 \pm 0.231$ & $0.561 \pm 0.029$ \\ [0.8ex]
      \hline
      Conc 3 & $35.274 \pm 1.210$ & $0.059 \pm 0.003$ & $81.363 \pm 0.722$ & $0.463 \pm 0.016$\\ [0.8ex]
      \hline
    \end{tabular}
    \caption{Time dependent autocorrelation functions (ACF) of inferred active forces are fitted by a double exponential function of the form, $f(t) = A_1 \exp(-t/\tau_1) + A_2 \exp(-t/\tau_2)$. The respective values of the corresponding fitting parameters along with errors are tabulated here. }
    \label{tab:acf_fit_param}
\end{table*}
\end{widetext}


\section{\label{sec:level5}Conclusions:}
In summary, we have numerically and experimentally demonstrated that the fluctuating active force on a spatially confined passive particle in an active fluid can be measured with high statistical accuracy by isolating the intertwined thermally induced random force. In doing so, this study has also revealed the true statistical characteristics of the active force, which is non-Gaussian and exhibits multi-exponential correlations. When the probability distribution function of the active force is modelled as a weighted sum of Gaussian and exponential distributions, the corresponding auto-correlation function fits a double-exponential form, with the slower time scale comparable to that of the active components. Additionally, a smaller time scale may arise from steric interactions among the constituents. It is important to note that our demonstrated method of inferring active force trajectory is independent of the nature of the force. Moreover, it requires minimal information about the system and is based on explicitly measurable quantities. Even though the consequences of decoding the force fluctuations in detail for such active systems have immense potential regarding the estimation of path-specific thermodynamic quantities~\cite{das2025localizing,huang2025entropy}, we feel that confirming those by thorough research is beyond the scope of the present work.

Finally, this approach can be extended to systems involving multiple passive particles, enabling the measurement of thermodynamic properties~\cite{dabelow2019irreversibility} with significantly enhanced precision. A very useful extension would also be for viscoelastic fluids, where the presence of memory would definitely lead to complications and unique features~\cite{narinder2019active}, but a successful strategy would be very useful for living matter systems which are inherently viscoelastic, and where activity is widely present~\cite{nakul2023studying,muenker2024accessing}. Additionally, inertial systems with complex interactions displaying intricate and exotic phenomena~\cite{pahi2025simultaneous} are the other avenues to apply this method to model their dynamics appropriately.  We are working in these areas, and hope to report interesting results in the future.  


\appendix

\section{\label{filter_accuracy}Filter accuracy: Dependence on the sampling time interval and total sampling time}
\begin{figure}[h!]
  \includegraphics[width=0.5\textwidth]{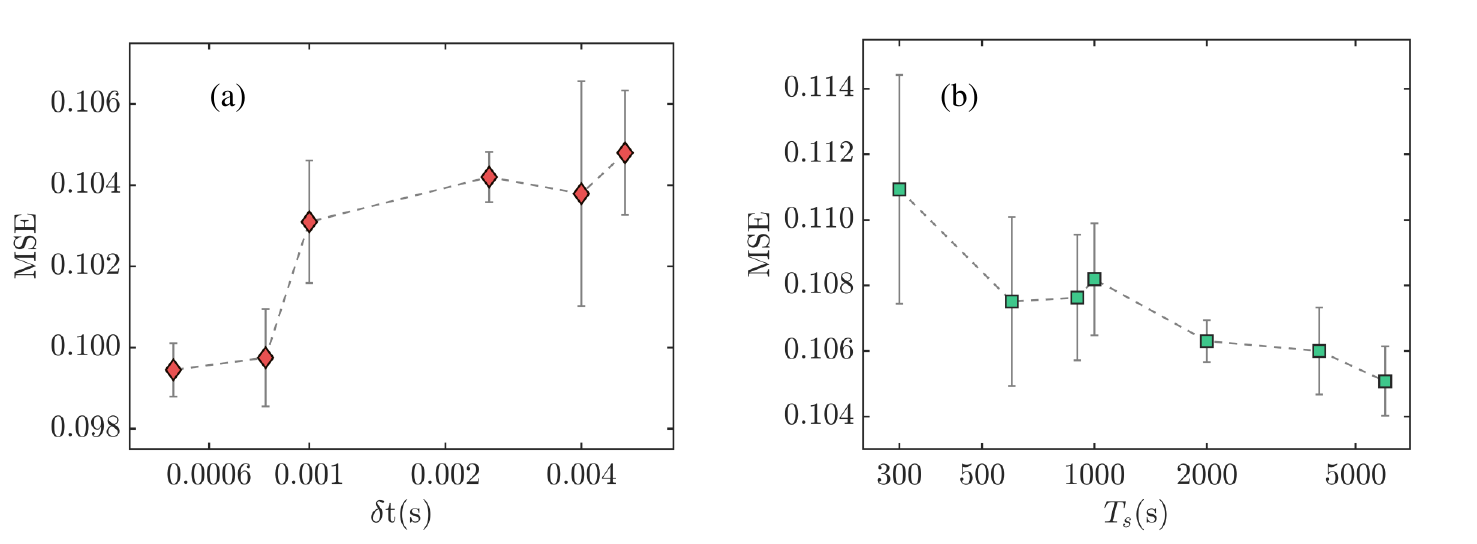}  
  \caption{Mean square error (MSE) plotted for various sampling rate ($\delta$t) and total time ($T_{s}$). Error bars represent standard deviations averaged over multiple independent inputs of $\delta$t and t. }
  \label{mse}
\end{figure}

The mean square error (MSE), as described in the main text (see \textit{section}~\ref{sec:level3}), is observed as a function of varying input parameters, such as sampling time step or rate ($\delta$t) and total sampling time ($T_{s}$). In Fig.~\ref{mse}(a), with a fixed sampling time ($T_{s}=1000$ s), MSE increases by around $4~\%$ with the increase of $\delta$t over one order of magnitude. Understandably, when $\delta$t is well below the characteristic timescale ($\approx$ 0.006 s for this case) of the system, the accuracy converges towards the maximum, and the MSE tends to the minimum.

Further, with the total sampling time and for a fixed $\delta t (=$ 0.005 s), in Fig.~\ref{mse}(b), reasonably, MSE declines for ascending values of `$T_{s}$'. However, the improvement in accuracy for this case is around $6~\%$ for an increase of $t$ from $300$ to $5000$ s. Further increase in $t$ does not seem to increase the accuracy any significantly.

\begin{figure*}[t]
  \includegraphics[width=1\textwidth]{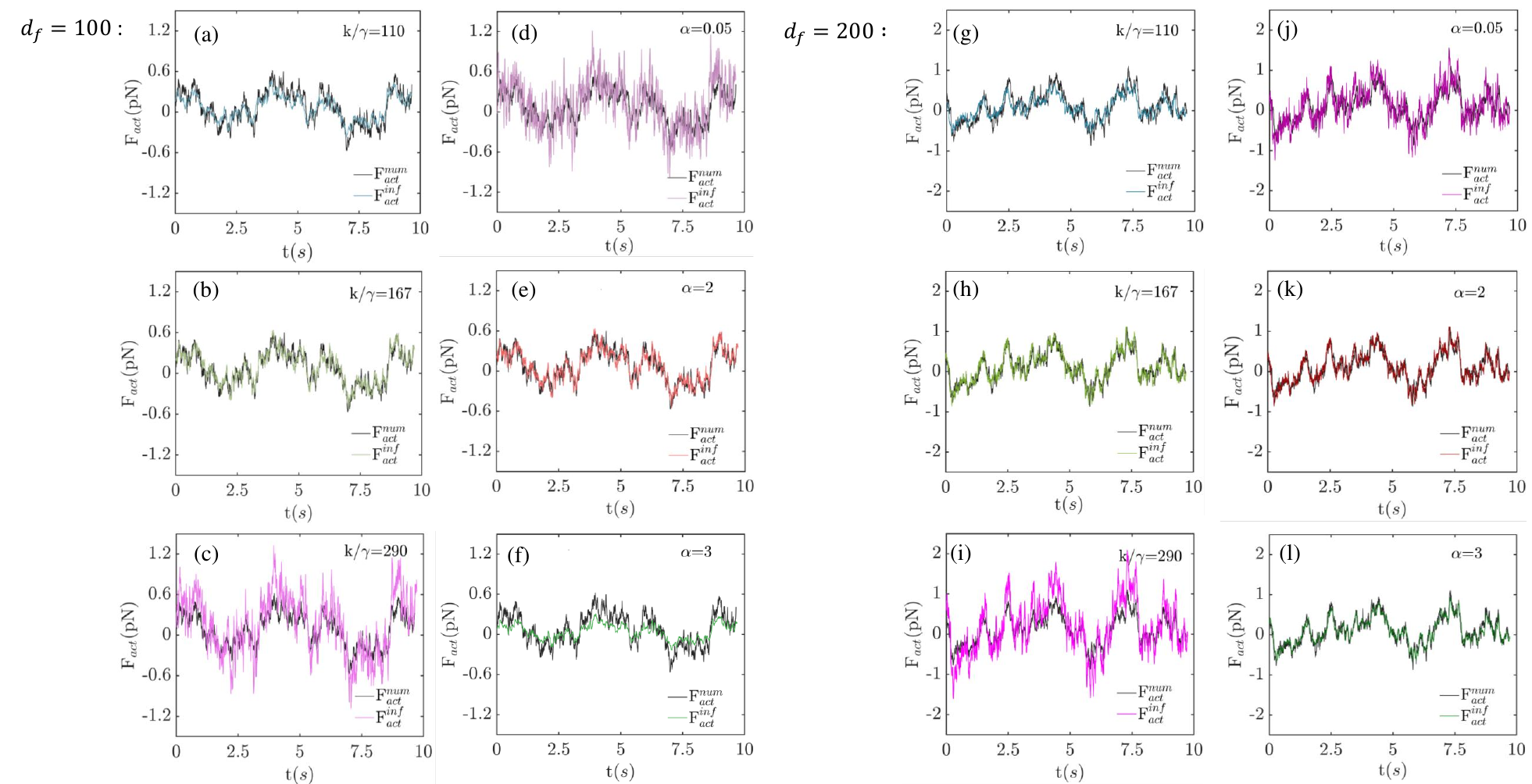}  
  \caption{Panels (a)–(f) display results for \(d_f = 100\), while panels (g)–(l) correspond to \(d_f = 200\). In panels (a)–(c) and (g)–(i), we fix \(\alpha = 2\) and vary the ratio \(k/\gamma\). Conversely, in panels (d)–(f) and (j)–(l), \(k/\gamma\) is maintained at \(167\,\mathrm{s}^{-1}\) while \(\alpha\) is scanned as the input parameter for the Wiener filter. Notably, the inference error is minimized at \(\alpha = 2\) and \(k/\gamma = 167\,\mathrm{s}^{-1}\).}
  \label{mse2}
\end{figure*}

\section{\label{filter_optimization} Optimization based on the system time scale and the thermal noise strength:}
\begin{figure}[h!]
  \includegraphics[width=0.48\textwidth]{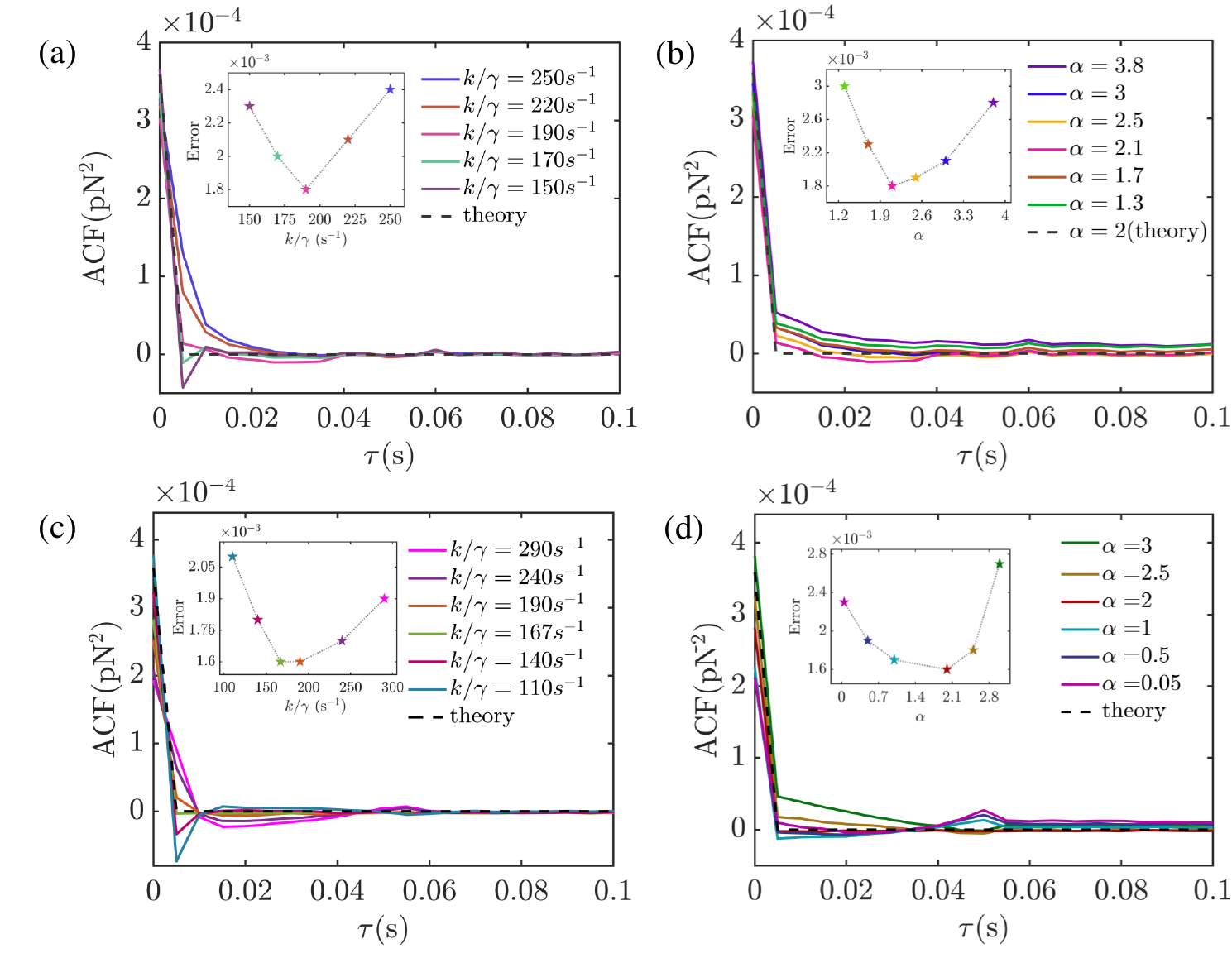}  
  \caption{(a)-(b) Experimental (for the highest active strength; i.e., Conc 3): The variation of ACF of thermal forces from its anticipated theoretical value minimizes for a particular measure of $\alpha$ and $k/\gamma$. In (a) $\alpha$ is fixed at 2.1 and in (b) $k/\gamma$ is fixed at $190 ~\text s^{-1}$. (c)-(d) Numerical: The deviation of ACFs of the inferred thermal forces are observed by varying the parameters $\alpha$ and $k/\gamma$ (for highest active strength, i.e., $d_f = 200$). In (c) $\alpha$ = 2 and in (d)   $k/\gamma$ = $167 ~\text s^{-1}$ are fixed. The corresponding errors for (a)-(d) are shown in the insets.}
  \label{alpha}
\end{figure}

Theoretically, one would expect that the trapped particle inside the active bacterial bath will follow the \textit{Langevin} equation - which can be discretized for the filtering purpose - as: $x_{t+dt} =x_t - (k/\gamma)x_t dt + (F^{act}_t/\gamma) dt + \sqrt{2k_B Tdt/\gamma}\ \eta_t$ with $\eta_t\sim \mathcal{N}(0,1)$. However,
due to the presence of the bacteria in the bath, the time scale of the system, i.e., $\gamma /k$ can be modified, and it should also be reflected in the strength of the thermal noise. Nonetheless, the statistical nature of the noise should remain unchanged with the subtle variations in the noise strength. Therefore, we write the variance of the noise as $\alpha k_BT/\gamma$ (where theoretically $\alpha$ should be equal to $2$
), and iteratively vary the inverse timescale $k/\gamma $ and $\alpha$ to obtain the noise with properties the same as known previously. For example, in Fig.~\ref{alpha}, we show how the autocorrelation function of random forces ($\xi_{theory} \sim \mathcal{N}(0,2k_B T \gamma)$) varies with these two parameters with respect to the reference known properties. Clearly, for typical experimental data, the deviation of the ACF from the theoretically expected nature is the minimum for certain values of $\alpha$ and $ k/\gamma$ (for the highest concentration, i.e., Conc 3, these are $\alpha \approx 2.1$ and $k/\gamma \approx 190 ~\text s^{-1}$) as shown in Fig.~\ref{alpha}(a)-(b). We also show this for numerically simulated trajectories in Fig.~\ref{alpha}(c)-(d). Note that the ACFs of the inferred random forces from the numerical trajectories are with the least error for $\alpha \approx 2$ and $k/\gamma \approx 167 ~\text s^{-1}$ as these are the actual values of the parameters used to generate the trajectories. The corresponding \textit{Error}s for each parameter sweep are plotted as the insets of Fig.~\ref{alpha}(a)-(d), and it is defined as, \textit{Error} = $|$ACF(\textit{Inferred random force}) - ACF(\textit{theory})$|$/($2k_BT\gamma$).  

In the Fig.~\ref{mse2}, we highlight the real-time correspondence between the input trajectories of the active force and their inferred values for two distinct \(d_f\) settings (100 and 200). An examination of the parameter space defined by \(k/\gamma\) and \(\alpha\) reveals that the inference error is minimized -- indicating an optimal match at \(k/\gamma = 167\ \text{s}^{-1}\) and \(\alpha = 2\). These optimal parameters coincide precisely with those used to generate the probe’s positional fluctuation trajectories in the simulations. Consequently, this analysis confirms that our method, based on minimizing the mean-squared error (MSE) of the thermal noise autocorrelation function, is both reliable and effective.

\section{\label{ecoli_culture}Sample preparation: Ecoli culture}
\textit{Motile E.Coli} (DH5a) cells are used as active particles. To facilitate differential count and imaging, \textit{E. coli} cells were transformed with the pTurboGFP-B plasmid (kindly provided by Dr. P. P. Dutta, IISER Kolkata) by the heat shock method \cite{gupta2022combined}. Positive transformants were selected on the basis of antibiotic selection, green fluorescence, and colony PCR. Next, GFP expressing-\textit{E.coli} were grown in Luria-Bertani (LB) medium (HiMedia, India) at 37$^\circ \text C$, supplemented with ampicillin (100 $\text \textmu \text g/\text {mL}$). The density of the bacterial culture was adjusted to an optical density (OD) of ~0.7 at 600 nm, corresponding to approximately $1 \times 10^7$ CFU/mL, as determined using a spectrophotometer. A 1 mL aliquot of the bacterial culture was pelleted by centrifugation at 4000 rpm for 7 minutes. The resulting pellet was resuspended in 1 mL of distilled water. The bacterial concentration was further adjusted according to the desired density.

\section{\label{ap:simulation}Simulation Methodology: Dynamics of passive probe in the bath of ABPs}

We performed simulations to study the behaviour of active Brownian particles (ABPs) interacting with a single passive particle in a two-dimensional periodic domain of size \(L = 20 \, \mu\text{m}\). The dynamics are governed by a set of stochastic differential equations (SDEs) that account for self-propulsion, interparticle interactions, confinement, and thermal fluctuations.
\subsection{Governing Equations}
Active Particles:  
The position \(\mathbf{r}_i = (x_i, y_i)\) and orientation \(\theta_i\) of the \(i\)-th ABP evolve as
\begin{equation}
 d\mathbf{r}_i = \left[v_0 \mathbf{e}(\theta_i) + \frac{1}{\gamma_\mathrm{active}} \mathbf{F}_i^\mathrm{int}\right] dt + \sqrt{2D_\mathrm{active}}\, d\mathbf{W}_i(t),   
\end{equation}
\begin{equation}
d\theta_i = \sqrt{2D_\mathrm{rot} }\, dW_i^\theta(t),
\end{equation}
where,
\(v_0\) is the self-propulsion speed,
\(\mathbf{e}(\theta_i) = (\cos\theta_i, \sin\theta_i)\) is the unit vector in the direction of motion,
\(\gamma_\mathrm{active} = 6\pi\eta r_\mathrm{active}\) is the friction coefficient for ABPs,
\(D_\mathrm{active} = k_BT/\gamma_\mathrm{active}\) is the translational diffusion coefficient,
\(D_\mathrm{rot} = k_BT/(8\pi\eta r_\mathrm{active}^3)\) is the rotational diffusion coefficient,
 \(\mathbf{F}_i^\mathrm{int}\) represents the net interaction force on particle \(i\) (including both active-active steric repulsion and active-passive interactions), \(d\mathbf{W}_i(t)\) and \(dW_i^\theta(t)\) are independent Wiener increments.

Passive Particle:  
The passive particle, located at \(\mathbf{r}_p\), experiences forces due to interactions with ABPs and a harmonic trapping force. Its dynamics are given by

\begin{equation}
 d\mathbf{r}_p = \frac{1}{\gamma_\mathrm{passive}} \left[\sum_{i=1}^{N}\mathbf{F}_{i\rightarrow p} + \mathbf{f}_\mathrm{trap}\right] dt + \sqrt{2D_\mathrm{passive}}\, d\mathbf{W}_p(t),   
\end{equation}
with:
\(\gamma_\mathrm{passive} = 6\pi\eta r_\mathrm{passive}\) and \(D_\mathrm{passive} = k_BT/\gamma_\mathrm{passive}\),
\(\mathbf{F}_{i\rightarrow p}\) representing the steric interaction force between the \(i\)-th active and the passive particle,
\(\mathbf{f}_\mathrm{trap} = -k_\mathrm{trap}(\mathbf{r}_p - L/2)\) is the harmonic trap force (with \(k_\mathrm{trap} = 7\times10^{-6}\) N/m),
\(d\mathbf{W}_p(t)\) is a Wiener increment of thermal fluctuations corresponding to the passive particle.

\subsection{Simulation Details}
The SDEs were discretized using the Euler–Maruyama method with timestep \(\Delta t = 0.0005\) s over a simulation duration of 200 s. Inter-particle forces were computed as follows:

1. Active-Active Interactions: 
   Steric repulsion was modeled for pairs of ABPs overlapping within a distance of \(2r_\mathrm{active}\) using a force proportional to the overlap. These forces were symmetrically distributed between particle pairs to satisfy Newton’s third law.

2. Active-Passive Interactions:  
   A similar steric repulsion was employed for ABP and passive particle pairs within a distance of \(r_\mathrm{active} + r_\mathrm{passive}\). The resulting forces are applied oppositely to the ABPs and the passive particle.

3. Trap Force:  
   The passive particle is confined by a harmonic potential centred at \( (L/2, L/2) \).

Positions and orientations are updated in each time step by combining the deterministic contributions from forces with stochastic contributions from Gaussian noise, which are scaled according to the relevant diffusion coefficients.

\subsection{Data Collection and Analysis}
Throughout the simulation, trajectories, forces, and an array of observables were meticulously recorded to facilitate thorough post-processing and analysis, thereby elucidating the efficacy of the filter as detailed in this study.

\begin{acknowledgments}
The work is supported by IISER Kolkata, and the Science and Engineering Research Board (SERB), Department of Science and Technology, Government of India, through the research grant CRG/2022/002417. AM acknowledges IISER Kolkata IPh.D fellowship for financial support throughout this work. SP is thankful to Kandi Raj College for its support. BD is thankful to the Ministry Of Education of Government of India for financial support through the Prime Minister’s Research Fellowship (PMRF) grant.
\end{acknowledgments}

\nocite{*}


\begin{thebibliography}{61}%
\makeatletter
\providecommand \@ifxundefined [1]{%
 \@ifx{#1\undefined}
}%
\providecommand \@ifnum [1]{%
 \ifnum #1\expandafter \@firstoftwo
 \else \expandafter \@secondoftwo
 \fi
}%
\providecommand \@ifx [1]{%
 \ifx #1\expandafter \@firstoftwo
 \else \expandafter \@secondoftwo
 \fi
}%
\providecommand \natexlab [1]{#1}%
\providecommand \enquote  [1]{``#1''}%
\providecommand \bibnamefont  [1]{#1}%
\providecommand \bibfnamefont [1]{#1}%
\providecommand \citenamefont [1]{#1}%
\providecommand \href@noop [0]{\@secondoftwo}%
\providecommand \href [0]{\begingroup \@sanitize@url \@href}%
\providecommand \@href[1]{\@@startlink{#1}\@@href}%
\providecommand \@@href[1]{\endgroup#1\@@endlink}%
\providecommand \@sanitize@url [0]{\catcode `\\12\catcode `\$12\catcode `\&12\catcode `\#12\catcode `\^12\catcode `\_12\catcode `\%12\relax}%
\providecommand \@@startlink[1]{}%
\providecommand \@@endlink[0]{}%
\providecommand \url  [0]{\begingroup\@sanitize@url \@url }%
\providecommand \@url [1]{\endgroup\@href {#1}{\urlprefix }}%
\providecommand \urlprefix  [0]{URL }%
\providecommand \Eprint [0]{\href }%
\providecommand \doibase [0]{https://doi.org/}%
\providecommand \selectlanguage [0]{\@gobble}%
\providecommand \bibinfo  [0]{\@secondoftwo}%
\providecommand \bibfield  [0]{\@secondoftwo}%
\providecommand \translation [1]{[#1]}%
\providecommand \BibitemOpen [0]{}%
\providecommand \bibitemStop [0]{}%
\providecommand \bibitemNoStop [0]{.\EOS\space}%
\providecommand \EOS [0]{\spacefactor3000\relax}%
\providecommand \BibitemShut  [1]{\csname bibitem#1\endcsname}%
\let\auto@bib@innerbib\@empty
\bibitem [{\citenamefont {Fodor}\ \emph {et~al.}(2016)\citenamefont {Fodor}, \citenamefont {Nardini}, \citenamefont {Cates}, \citenamefont {Tailleur}, \citenamefont {Visco},\ and\ \citenamefont {van Wijland}}]{fodor2016far}%
  \BibitemOpen
  \bibfield  {author} {\bibinfo {author} {\bibfnamefont {E.}~\bibnamefont {Fodor}}, \bibinfo {author} {\bibfnamefont {C.}~\bibnamefont {Nardini}}, \bibinfo {author} {\bibfnamefont {M.~E.}\ \bibnamefont {Cates}}, \bibinfo {author} {\bibfnamefont {J.}~\bibnamefont {Tailleur}}, \bibinfo {author} {\bibfnamefont {P.}~\bibnamefont {Visco}},\ and\ \bibinfo {author} {\bibfnamefont {F.}~\bibnamefont {van Wijland}},\ }\bibfield  {title} {\bibinfo {title} {How far from equilibrium is active matter?},\ }\href {https://doi.org/10.1103/PhysRevLett.117.038103} {\bibfield  {journal} {\bibinfo  {journal} {Physical Review Letters}\ }\textbf {\bibinfo {volume} {117}},\ \bibinfo {pages} {038103} (\bibinfo {year} {2016})}\BibitemShut {NoStop}%
\bibitem [{\citenamefont {Ramaswamy}(2017)}]{ramaswamy2017active}%
  \BibitemOpen
  \bibfield  {author} {\bibinfo {author} {\bibfnamefont {S.}~\bibnamefont {Ramaswamy}},\ }\bibfield  {title} {\bibinfo {title} {Active matter},\ }\href {https://doi.org/10.1088/1742-5468/aa6bc5} {\bibfield  {journal} {\bibinfo  {journal} {Journal of Statistical Mechanics: Theory and Experiment}\ }\textbf {\bibinfo {volume} {2017}},\ \bibinfo {pages} {054002} (\bibinfo {year} {2017})}\BibitemShut {NoStop}%
\bibitem [{\citenamefont {Vrugt}\ and\ \citenamefont {Wittkowski}(2024)}]{vrugt2024review}%
  \BibitemOpen
  \bibfield  {author} {\bibinfo {author} {\bibfnamefont {M.~t.}\ \bibnamefont {Vrugt}}\ and\ \bibinfo {author} {\bibfnamefont {R.}~\bibnamefont {Wittkowski}},\ }\bibfield  {title} {\bibinfo {title} {A review of active matter reviews},\ }\href {https://doi.org/10.48550/arXiv.2405.15751} {\bibfield  {journal} {\bibinfo  {journal} {arXiv preprint arXiv:2405.15751}\ } (\bibinfo {year} {2024})}\BibitemShut {NoStop}%
\bibitem [{\citenamefont {Bechinger}\ \emph {et~al.}(2016)\citenamefont {Bechinger}, \citenamefont {Di~Leonardo}, \citenamefont {L{\"o}wen}, \citenamefont {Reichhardt}, \citenamefont {Volpe},\ and\ \citenamefont {Volpe}}]{bechinger2016active}%
  \BibitemOpen
  \bibfield  {author} {\bibinfo {author} {\bibfnamefont {C.}~\bibnamefont {Bechinger}}, \bibinfo {author} {\bibfnamefont {R.}~\bibnamefont {Di~Leonardo}}, \bibinfo {author} {\bibfnamefont {H.}~\bibnamefont {L{\"o}wen}}, \bibinfo {author} {\bibfnamefont {C.}~\bibnamefont {Reichhardt}}, \bibinfo {author} {\bibfnamefont {G.}~\bibnamefont {Volpe}},\ and\ \bibinfo {author} {\bibfnamefont {G.}~\bibnamefont {Volpe}},\ }\bibfield  {title} {\bibinfo {title} {Active particles in complex and crowded environments},\ }\href {https://doi.org/10.1103/RevModPhys.88.045006} {\bibfield  {journal} {\bibinfo  {journal} {Reviews of Modern Physics}\ }\textbf {\bibinfo {volume} {88}},\ \bibinfo {pages} {045006} (\bibinfo {year} {2016})}\BibitemShut {NoStop}%
\bibitem [{\citenamefont {Bowick}\ \emph {et~al.}(2022)\citenamefont {Bowick}, \citenamefont {Fakhri}, \citenamefont {Marchetti},\ and\ \citenamefont {Ramaswamy}}]{bowick2022symmetry}%
  \BibitemOpen
  \bibfield  {author} {\bibinfo {author} {\bibfnamefont {M.~J.}\ \bibnamefont {Bowick}}, \bibinfo {author} {\bibfnamefont {N.}~\bibnamefont {Fakhri}}, \bibinfo {author} {\bibfnamefont {M.~C.}\ \bibnamefont {Marchetti}},\ and\ \bibinfo {author} {\bibfnamefont {S.}~\bibnamefont {Ramaswamy}},\ }\bibfield  {title} {\bibinfo {title} {Symmetry, thermodynamics, and topology in active matter},\ }\href {https://doi.org/10.1103/PhysRevX.12.010501} {\bibfield  {journal} {\bibinfo  {journal} {Physical Review X}\ }\textbf {\bibinfo {volume} {12}},\ \bibinfo {pages} {010501} (\bibinfo {year} {2022})}\BibitemShut {NoStop}%
\bibitem [{\citenamefont {Wu}\ and\ \citenamefont {Libchaber}(2000)}]{wu2000particle}%
  \BibitemOpen
  \bibfield  {author} {\bibinfo {author} {\bibfnamefont {X.-L.}\ \bibnamefont {Wu}}\ and\ \bibinfo {author} {\bibfnamefont {A.}~\bibnamefont {Libchaber}},\ }\bibfield  {title} {\bibinfo {title} {Particle diffusion in a quasi-two-dimensional bacterial bath},\ }\href {https://link.aps.org/doi/10.1103/PhysRevLett.84.3017} {\bibfield  {journal} {\bibinfo  {journal} {Physical Review Letters}\ }\textbf {\bibinfo {volume} {84}},\ \bibinfo {pages} {3017} (\bibinfo {year} {2000})}\BibitemShut {NoStop}%
\bibitem [{\citenamefont {Leptos}\ \emph {et~al.}(2009)\citenamefont {Leptos}, \citenamefont {Guasto}, \citenamefont {Gollub}, \citenamefont {Pesci},\ and\ \citenamefont {Goldstein}}]{leptos2009dynamics}%
  \BibitemOpen
  \bibfield  {author} {\bibinfo {author} {\bibfnamefont {K.~C.}\ \bibnamefont {Leptos}}, \bibinfo {author} {\bibfnamefont {J.~S.}\ \bibnamefont {Guasto}}, \bibinfo {author} {\bibfnamefont {J.~P.}\ \bibnamefont {Gollub}}, \bibinfo {author} {\bibfnamefont {A.~I.}\ \bibnamefont {Pesci}},\ and\ \bibinfo {author} {\bibfnamefont {R.~E.}\ \bibnamefont {Goldstein}},\ }\bibfield  {title} {\bibinfo {title} {Dynamics of enhanced tracer diffusion in suspensions of swimming eukaryotic microorganisms},\ }\href {http://dx.doi.org/10.1103/PhysRevLett.103.198103} {\bibfield  {journal} {\bibinfo  {journal} {Physical Review Letters}\ }\textbf {\bibinfo {volume} {103}},\ \bibinfo {pages} {198103} (\bibinfo {year} {2009})}\BibitemShut {NoStop}%
\bibitem [{\citenamefont {Argun}\ \emph {et~al.}(2016)\citenamefont {Argun}, \citenamefont {Moradi}, \citenamefont {Pin{\c{c}}e}, \citenamefont {Bagci}, \citenamefont {Imparato},\ and\ \citenamefont {Volpe}}]{argun2016non}%
  \BibitemOpen
  \bibfield  {author} {\bibinfo {author} {\bibfnamefont {A.}~\bibnamefont {Argun}}, \bibinfo {author} {\bibfnamefont {A.-R.}\ \bibnamefont {Moradi}}, \bibinfo {author} {\bibfnamefont {E.}~\bibnamefont {Pin{\c{c}}e}}, \bibinfo {author} {\bibfnamefont {G.~B.}\ \bibnamefont {Bagci}}, \bibinfo {author} {\bibfnamefont {A.}~\bibnamefont {Imparato}},\ and\ \bibinfo {author} {\bibfnamefont {G.}~\bibnamefont {Volpe}},\ }\bibfield  {title} {\bibinfo {title} {Non-boltzmann stationary distributions and nonequilibrium relations in active baths},\ }\href {https://doi.org/10.1103/PhysRevE.94.062150} {\bibfield  {journal} {\bibinfo  {journal} {Physical Review E}\ }\textbf {\bibinfo {volume} {94}},\ \bibinfo {pages} {062150} (\bibinfo {year} {2016})}\BibitemShut {NoStop}%
\bibitem [{\citenamefont {Ortlieb}\ \emph {et~al.}(2019)\citenamefont {Ortlieb}, \citenamefont {Rafa\"{\i}}, \citenamefont {Peyla}, \citenamefont {Wagner},\ and\ \citenamefont {John}}]{ortileb2019statistics}%
  \BibitemOpen
  \bibfield  {author} {\bibinfo {author} {\bibfnamefont {L.}~\bibnamefont {Ortlieb}}, \bibinfo {author} {\bibfnamefont {S.}~\bibnamefont {Rafa\"{\i}}}, \bibinfo {author} {\bibfnamefont {P.}~\bibnamefont {Peyla}}, \bibinfo {author} {\bibfnamefont {C.}~\bibnamefont {Wagner}},\ and\ \bibinfo {author} {\bibfnamefont {T.}~\bibnamefont {John}},\ }\bibfield  {title} {\bibinfo {title} {Statistics of colloidal suspensions stirred by microswimmers},\ }\href {https://doi.org/10.1103/PhysRevLett.122.148101} {\bibfield  {journal} {\bibinfo  {journal} {Physical Review Letters}\ }\textbf {\bibinfo {volume} {122}},\ \bibinfo {pages} {148101} (\bibinfo {year} {2019})}\BibitemShut {NoStop}%
\bibitem [{\citenamefont {Maggi}\ \emph {et~al.}(2014)\citenamefont {Maggi}, \citenamefont {Paoluzzi}, \citenamefont {Pellicciotta}, \citenamefont {Lepore}, \citenamefont {Angelani},\ and\ \citenamefont {Di~Leonardo}}]{maggi2014generalized}%
  \BibitemOpen
  \bibfield  {author} {\bibinfo {author} {\bibfnamefont {C.}~\bibnamefont {Maggi}}, \bibinfo {author} {\bibfnamefont {M.}~\bibnamefont {Paoluzzi}}, \bibinfo {author} {\bibfnamefont {N.}~\bibnamefont {Pellicciotta}}, \bibinfo {author} {\bibfnamefont {A.}~\bibnamefont {Lepore}}, \bibinfo {author} {\bibfnamefont {L.}~\bibnamefont {Angelani}},\ and\ \bibinfo {author} {\bibfnamefont {R.}~\bibnamefont {Di~Leonardo}},\ }\bibfield  {title} {\bibinfo {title} {Generalized energy equipartition in harmonic oscillators driven by active baths},\ }\href {https://doi.org/10.1103/PhysRevLett.113.238303} {\bibfield  {journal} {\bibinfo  {journal} {Physical Review Letters}\ }\textbf {\bibinfo {volume} {113}},\ \bibinfo {pages} {238303} (\bibinfo {year} {2014})}\BibitemShut {NoStop}%
\bibitem [{\citenamefont {Dor}\ \emph {et~al.}(2022)\citenamefont {Dor}, \citenamefont {Kafri}, \citenamefont {Kardar},\ and\ \citenamefont {Tailleur}}]{dor2022passive}%
  \BibitemOpen
  \bibfield  {author} {\bibinfo {author} {\bibfnamefont {Y.~B.}\ \bibnamefont {Dor}}, \bibinfo {author} {\bibfnamefont {Y.}~\bibnamefont {Kafri}}, \bibinfo {author} {\bibfnamefont {M.}~\bibnamefont {Kardar}},\ and\ \bibinfo {author} {\bibfnamefont {J.}~\bibnamefont {Tailleur}},\ }\bibfield  {title} {\bibinfo {title} {Passive objects in confined active fluids: A localization transition},\ }\href {https://doi.org/10.1103/PhysRevE.106.044604} {\bibfield  {journal} {\bibinfo  {journal} {Phys. Rev. E}\ }\textbf {\bibinfo {volume} {106}},\ \bibinfo {pages} {044604} (\bibinfo {year} {2022})}\BibitemShut {NoStop}%
\bibitem [{\citenamefont {Reichert}\ and\ \citenamefont {Voigtmann}(2021)}]{reichert2021tracer}%
  \BibitemOpen
  \bibfield  {author} {\bibinfo {author} {\bibfnamefont {J.}~\bibnamefont {Reichert}}\ and\ \bibinfo {author} {\bibfnamefont {T.}~\bibnamefont {Voigtmann}},\ }\bibfield  {title} {\bibinfo {title} {Tracer dynamics in crowded active-particle suspensions},\ }\href {https://doi.org/10.1039/D1SM01092A} {\bibfield  {journal} {\bibinfo  {journal} {Soft Matter}\ }\textbf {\bibinfo {volume} {17}},\ \bibinfo {pages} {10492} (\bibinfo {year} {2021})}\BibitemShut {NoStop}%
\bibitem [{\citenamefont {Kushwaha}\ \emph {et~al.}(2023)\citenamefont {Kushwaha}, \citenamefont {Semwal}, \citenamefont {Maity}, \citenamefont {Mishra},\ and\ \citenamefont {Chikkadi}}]{kushwaha2023phase}%
  \BibitemOpen
  \bibfield  {author} {\bibinfo {author} {\bibfnamefont {P.}~\bibnamefont {Kushwaha}}, \bibinfo {author} {\bibfnamefont {V.}~\bibnamefont {Semwal}}, \bibinfo {author} {\bibfnamefont {S.}~\bibnamefont {Maity}}, \bibinfo {author} {\bibfnamefont {S.}~\bibnamefont {Mishra}},\ and\ \bibinfo {author} {\bibfnamefont {V.}~\bibnamefont {Chikkadi}},\ }\bibfield  {title} {\bibinfo {title} {Phase separation of passive particles in active liquids},\ }\href {https://doi.org/10.1103/PhysRevE.108.034603} {\bibfield  {journal} {\bibinfo  {journal} {Physical Review E}\ }\textbf {\bibinfo {volume} {108}},\ \bibinfo {pages} {034603} (\bibinfo {year} {2023})}\BibitemShut {NoStop}%
\bibitem [{\citenamefont {Singh}\ and\ \citenamefont {Chaudhuri}(2024)}]{singh2024anomalous}%
  \BibitemOpen
  \bibfield  {author} {\bibinfo {author} {\bibfnamefont {C.}~\bibnamefont {Singh}}\ and\ \bibinfo {author} {\bibfnamefont {A.}~\bibnamefont {Chaudhuri}},\ }\bibfield  {title} {\bibinfo {title} {Anomalous dynamics of a passive droplet in active turbulence},\ }\href {https://doi.org/10.1038/s41467-024-47727-1} {\bibfield  {journal} {\bibinfo  {journal} {Nature Communications}\ }\textbf {\bibinfo {volume} {15}},\ \bibinfo {pages} {3704} (\bibinfo {year} {2024})}\BibitemShut {NoStop}%
\bibitem [{\citenamefont {Caprini}\ \emph {et~al.}(2024)\citenamefont {Caprini}, \citenamefont {Ldov}, \citenamefont {Gupta}, \citenamefont {Ellenberg}, \citenamefont {Wittmann}, \citenamefont {L{\"o}wen},\ and\ \citenamefont {Scholz}}]{caprini2024emergent}%
  \BibitemOpen
  \bibfield  {author} {\bibinfo {author} {\bibfnamefont {L.}~\bibnamefont {Caprini}}, \bibinfo {author} {\bibfnamefont {A.}~\bibnamefont {Ldov}}, \bibinfo {author} {\bibfnamefont {R.~K.}\ \bibnamefont {Gupta}}, \bibinfo {author} {\bibfnamefont {H.}~\bibnamefont {Ellenberg}}, \bibinfo {author} {\bibfnamefont {R.}~\bibnamefont {Wittmann}}, \bibinfo {author} {\bibfnamefont {H.}~\bibnamefont {L{\"o}wen}},\ and\ \bibinfo {author} {\bibfnamefont {C.}~\bibnamefont {Scholz}},\ }\bibfield  {title} {\bibinfo {title} {Emergent memory from tapping collisions in active granular matter},\ }\href {https://doi.org/10.1038/s42005-024-01540-w} {\bibfield  {journal} {\bibinfo  {journal} {Communications Physics}\ }\textbf {\bibinfo {volume} {7}},\ \bibinfo {pages} {52} (\bibinfo {year} {2024})}\BibitemShut {NoStop}%
\bibitem [{\citenamefont {Lagarde}\ \emph {et~al.}(2020)\citenamefont {Lagarde}, \citenamefont {Dag{\`e}s}, \citenamefont {Nemoto}, \citenamefont {D{\'e}mery}, \citenamefont {Bartolo},\ and\ \citenamefont {Gibaud}}]{lagarde2020colloidal}%
  \BibitemOpen
  \bibfield  {author} {\bibinfo {author} {\bibfnamefont {A.}~\bibnamefont {Lagarde}}, \bibinfo {author} {\bibfnamefont {N.}~\bibnamefont {Dag{\`e}s}}, \bibinfo {author} {\bibfnamefont {T.}~\bibnamefont {Nemoto}}, \bibinfo {author} {\bibfnamefont {V.}~\bibnamefont {D{\'e}mery}}, \bibinfo {author} {\bibfnamefont {D.}~\bibnamefont {Bartolo}},\ and\ \bibinfo {author} {\bibfnamefont {T.}~\bibnamefont {Gibaud}},\ }\bibfield  {title} {\bibinfo {title} {Colloidal transport in bacteria suspensions: from bacteria collision to anomalous and enhanced diffusion},\ }\href {https://doi.org/10.1039/D0SM00309C} {\bibfield  {journal} {\bibinfo  {journal} {Soft Matter}\ }\textbf {\bibinfo {volume} {16}},\ \bibinfo {pages} {7503} (\bibinfo {year} {2020})}\BibitemShut {NoStop}%
\bibitem [{\citenamefont {Nordanger}\ \emph {et~al.}(2022)\citenamefont {Nordanger}, \citenamefont {Morozov},\ and\ \citenamefont {Stenhammar}}]{nordanger2022anisotropic}%
  \BibitemOpen
  \bibfield  {author} {\bibinfo {author} {\bibfnamefont {H.}~\bibnamefont {Nordanger}}, \bibinfo {author} {\bibfnamefont {A.}~\bibnamefont {Morozov}},\ and\ \bibinfo {author} {\bibfnamefont {J.}~\bibnamefont {Stenhammar}},\ }\bibfield  {title} {\bibinfo {title} {Anisotropic diffusion of ellipsoidal tracers in microswimmer suspensions},\ }\href {https://doi.org/10.1103/PhysRevFluids.7.013103} {\bibfield  {journal} {\bibinfo  {journal} {Phys. Rev. Fluids}\ }\textbf {\bibinfo {volume} {7}},\ \bibinfo {pages} {013103} (\bibinfo {year} {2022})}\BibitemShut {NoStop}%
\bibitem [{\citenamefont {Santra}(2023)}]{santra2023dynamical}%
  \BibitemOpen
  \bibfield  {author} {\bibinfo {author} {\bibfnamefont {I.}~\bibnamefont {Santra}},\ }\bibfield  {title} {\bibinfo {title} {Dynamical fluctuations of a tracer coupled to active and passive particles},\ }\href {https://iopscience.iop.org/article/10.1088/2632-072X/acbf1a} {\bibfield  {journal} {\bibinfo  {journal} {Journal of Physics: Complexity}\ }\textbf {\bibinfo {volume} {4}},\ \bibinfo {pages} {015013} (\bibinfo {year} {2023})}\BibitemShut {NoStop}%
\bibitem [{\citenamefont {Goswami}\ \emph {et~al.}(2024)\citenamefont {Goswami}, \citenamefont {Cherstvy}, \citenamefont {Godec},\ and\ \citenamefont {Metzler}}]{goswami2024anomalous}%
  \BibitemOpen
  \bibfield  {author} {\bibinfo {author} {\bibfnamefont {K.}~\bibnamefont {Goswami}}, \bibinfo {author} {\bibfnamefont {A.~G.}\ \bibnamefont {Cherstvy}}, \bibinfo {author} {\bibfnamefont {A.}~\bibnamefont {Godec}},\ and\ \bibinfo {author} {\bibfnamefont {R.}~\bibnamefont {Metzler}},\ }\bibfield  {title} {\bibinfo {title} {Anomalous diffusion of active brownian particles in responsive elastic gels: Nonergodicity, non-gaussianity, and distributions of trapping times},\ }\href {https://doi.org/10.1103/PhysRevE.110.044609} {\bibfield  {journal} {\bibinfo  {journal} {Phys. Rev. E}\ }\textbf {\bibinfo {volume} {110}},\ \bibinfo {pages} {044609} (\bibinfo {year} {2024})}\BibitemShut {NoStop}%
\bibitem [{\citenamefont {Di~Leonardo}\ \emph {et~al.}(2010)\citenamefont {Di~Leonardo}, \citenamefont {Angelani}, \citenamefont {Dell’Arciprete}, \citenamefont {Ruocco}, \citenamefont {Iebba}, \citenamefont {Schippa}, \citenamefont {Conte}, \citenamefont {Mecarini}, \citenamefont {De~Angelis},\ and\ \citenamefont {Di~Fabrizio}}]{di2010bacterial}%
  \BibitemOpen
  \bibfield  {author} {\bibinfo {author} {\bibfnamefont {R.}~\bibnamefont {Di~Leonardo}}, \bibinfo {author} {\bibfnamefont {L.}~\bibnamefont {Angelani}}, \bibinfo {author} {\bibfnamefont {D.}~\bibnamefont {Dell’Arciprete}}, \bibinfo {author} {\bibfnamefont {G.}~\bibnamefont {Ruocco}}, \bibinfo {author} {\bibfnamefont {V.}~\bibnamefont {Iebba}}, \bibinfo {author} {\bibfnamefont {S.}~\bibnamefont {Schippa}}, \bibinfo {author} {\bibfnamefont {M.~P.}\ \bibnamefont {Conte}}, \bibinfo {author} {\bibfnamefont {F.}~\bibnamefont {Mecarini}}, \bibinfo {author} {\bibfnamefont {F.}~\bibnamefont {De~Angelis}},\ and\ \bibinfo {author} {\bibfnamefont {E.}~\bibnamefont {Di~Fabrizio}},\ }\bibfield  {title} {\bibinfo {title} {Bacterial ratchet motors},\ }\href {https://www.pnas.org/doi/full/10.1073/pnas.0910426107} {\bibfield  {journal} {\bibinfo  {journal} {Proceedings of the National Academy of Sciences}\ }\textbf {\bibinfo {volume} {107}},\ \bibinfo {pages} {9541} (\bibinfo {year} {2010})}\BibitemShut {NoStop}%
\bibitem [{\citenamefont {Sokolov}\ \emph {et~al.}(2010)\citenamefont {Sokolov}, \citenamefont {Apodaca}, \citenamefont {Grzybowski},\ and\ \citenamefont {Aranson}}]{sokolov2010swimming}%
  \BibitemOpen
  \bibfield  {author} {\bibinfo {author} {\bibfnamefont {A.}~\bibnamefont {Sokolov}}, \bibinfo {author} {\bibfnamefont {M.~M.}\ \bibnamefont {Apodaca}}, \bibinfo {author} {\bibfnamefont {B.~A.}\ \bibnamefont {Grzybowski}},\ and\ \bibinfo {author} {\bibfnamefont {I.~S.}\ \bibnamefont {Aranson}},\ }\bibfield  {title} {\bibinfo {title} {Swimming bacteria power microscopic gears},\ }\href {https://doi.org/10.1073/pnas.0913015107} {\bibfield  {journal} {\bibinfo  {journal} {Proceedings of the National Academy of Sciences}\ }\textbf {\bibinfo {volume} {107}},\ \bibinfo {pages} {969} (\bibinfo {year} {2010})}\BibitemShut {NoStop}%
\bibitem [{\citenamefont {Zaeifi~Yamchi}\ and\ \citenamefont {Naji}(2017)}]{zaeifi2017effective}%
  \BibitemOpen
  \bibfield  {author} {\bibinfo {author} {\bibfnamefont {M.}~\bibnamefont {Zaeifi~Yamchi}}\ and\ \bibinfo {author} {\bibfnamefont {A.}~\bibnamefont {Naji}},\ }\bibfield  {title} {\bibinfo {title} {Effective interactions between inclusions in an active bath},\ }\href {https://doi.org/10.1063/1.5001505} {\bibfield  {journal} {\bibinfo  {journal} {The Journal of chemical physics}\ }\textbf {\bibinfo {volume} {147}} (\bibinfo {year} {2017})}\BibitemShut {NoStop}%
\bibitem [{\citenamefont {Baek}\ \emph {et~al.}(2018)\citenamefont {Baek}, \citenamefont {Solon}, \citenamefont {Xu}, \citenamefont {Nikola},\ and\ \citenamefont {Kafri}}]{baek2018generic}%
  \BibitemOpen
  \bibfield  {author} {\bibinfo {author} {\bibfnamefont {Y.}~\bibnamefont {Baek}}, \bibinfo {author} {\bibfnamefont {A.~P.}\ \bibnamefont {Solon}}, \bibinfo {author} {\bibfnamefont {X.}~\bibnamefont {Xu}}, \bibinfo {author} {\bibfnamefont {N.}~\bibnamefont {Nikola}},\ and\ \bibinfo {author} {\bibfnamefont {Y.}~\bibnamefont {Kafri}},\ }\bibfield  {title} {\bibinfo {title} {Generic long-range interactions between passive bodies in an active fluid},\ }\href {https://doi.org/10.1103/PhysRevLett.120.058002} {\bibfield  {journal} {\bibinfo  {journal} {Phys. Rev. Lett.}\ }\textbf {\bibinfo {volume} {120}},\ \bibinfo {pages} {058002} (\bibinfo {year} {2018})}\BibitemShut {NoStop}%
\bibitem [{\citenamefont {Hernandez-Ortiz}\ \emph {et~al.}(2005)\citenamefont {Hernandez-Ortiz}, \citenamefont {Stoltz},\ and\ \citenamefont {Graham}}]{ortiz2005transport}%
  \BibitemOpen
  \bibfield  {author} {\bibinfo {author} {\bibfnamefont {J.~P.}\ \bibnamefont {Hernandez-Ortiz}}, \bibinfo {author} {\bibfnamefont {C.~G.}\ \bibnamefont {Stoltz}},\ and\ \bibinfo {author} {\bibfnamefont {M.~D.}\ \bibnamefont {Graham}},\ }\bibfield  {title} {\bibinfo {title} {Transport and collective dynamics in suspensions of confined swimming particles},\ }\href {https://doi.org/10.1103/PhysRevLett.95.204501} {\bibfield  {journal} {\bibinfo  {journal} {Phys. Rev. Lett.}\ }\textbf {\bibinfo {volume} {95}},\ \bibinfo {pages} {204501} (\bibinfo {year} {2005})}\BibitemShut {NoStop}%
\bibitem [{\citenamefont {Angelani}\ \emph {et~al.}(2011)\citenamefont {Angelani}, \citenamefont {Maggi}, \citenamefont {Bernardini}, \citenamefont {Rizzo},\ and\ \citenamefont {Di~Leonardo}}]{angelani2011effective}%
  \BibitemOpen
  \bibfield  {author} {\bibinfo {author} {\bibfnamefont {L.}~\bibnamefont {Angelani}}, \bibinfo {author} {\bibfnamefont {C.}~\bibnamefont {Maggi}}, \bibinfo {author} {\bibfnamefont {M.~L.}\ \bibnamefont {Bernardini}}, \bibinfo {author} {\bibfnamefont {A.}~\bibnamefont {Rizzo}},\ and\ \bibinfo {author} {\bibfnamefont {R.}~\bibnamefont {Di~Leonardo}},\ }\bibfield  {title} {\bibinfo {title} {Effective interactions between colloidal particles suspended in a bath of swimming cells},\ }\href {https://doi.org/10.1103/PhysRevLett.107.138302} {\bibfield  {journal} {\bibinfo  {journal} {Phys. Rev. Lett.}\ }\textbf {\bibinfo {volume} {107}},\ \bibinfo {pages} {138302} (\bibinfo {year} {2011})}\BibitemShut {NoStop}%
\bibitem [{\citenamefont {Paul}\ \emph {et~al.}(2022)\citenamefont {Paul}, \citenamefont {Jayaram}, \citenamefont {Narinder}, \citenamefont {Speck},\ and\ \citenamefont {Bechinger}}]{paul2022force}%
  \BibitemOpen
  \bibfield  {author} {\bibinfo {author} {\bibfnamefont {S.}~\bibnamefont {Paul}}, \bibinfo {author} {\bibfnamefont {A.}~\bibnamefont {Jayaram}}, \bibinfo {author} {\bibfnamefont {N.}~\bibnamefont {Narinder}}, \bibinfo {author} {\bibfnamefont {T.}~\bibnamefont {Speck}},\ and\ \bibinfo {author} {\bibfnamefont {C.}~\bibnamefont {Bechinger}},\ }\bibfield  {title} {\bibinfo {title} {Force generation in confined active fluids: The role of microstructure},\ }\href {https://doi.org/10.1103/PhysRevLett.129.058001} {\bibfield  {journal} {\bibinfo  {journal} {Physical Review Letters}\ }\textbf {\bibinfo {volume} {129}},\ \bibinfo {pages} {058001} (\bibinfo {year} {2022})}\BibitemShut {NoStop}%
\bibitem [{\citenamefont {Jayaram}\ and\ \citenamefont {Speck}(2023)}]{jayaram2023effective}%
  \BibitemOpen
  \bibfield  {author} {\bibinfo {author} {\bibfnamefont {A.}~\bibnamefont {Jayaram}}\ and\ \bibinfo {author} {\bibfnamefont {T.}~\bibnamefont {Speck}},\ }\bibfield  {title} {\bibinfo {title} {Effective dynamics and fluctuations of a trapped probe moving in a fluid of active hard discs (a)},\ }\href {https://doi.org/10.1209/0295-5075/acdf1a} {\bibfield  {journal} {\bibinfo  {journal} {Europhysics Letters}\ }\textbf {\bibinfo {volume} {143}},\ \bibinfo {pages} {17005} (\bibinfo {year} {2023})}\BibitemShut {NoStop}%
\bibitem [{\citenamefont {Shea}\ \emph {et~al.}(2024)\citenamefont {Shea}, \citenamefont {Jung},\ and\ \citenamefont {Schmid}}]{shea2024force}%
  \BibitemOpen
  \bibfield  {author} {\bibinfo {author} {\bibfnamefont {J.}~\bibnamefont {Shea}}, \bibinfo {author} {\bibfnamefont {G.}~\bibnamefont {Jung}},\ and\ \bibinfo {author} {\bibfnamefont {F.}~\bibnamefont {Schmid}},\ }\bibfield  {title} {\bibinfo {title} {Force renormalization for probes immersed in an active bath},\ }\href {https://doi.org/10.1039/D3SM01387A} {\bibfield  {journal} {\bibinfo  {journal} {Soft Matter}\ }\textbf {\bibinfo {volume} {20}},\ \bibinfo {pages} {1767} (\bibinfo {year} {2024})}\BibitemShut {NoStop}%
\bibitem [{\citenamefont {Solon}\ \emph {et~al.}(2015)\citenamefont {Solon}, \citenamefont {Fily}, \citenamefont {Baskaran}, \citenamefont {Cates}, \citenamefont {Kafri}, \citenamefont {Kardar},\ and\ \citenamefont {Tailleur}}]{solon2015pressure}%
  \BibitemOpen
  \bibfield  {author} {\bibinfo {author} {\bibfnamefont {A.~P.}\ \bibnamefont {Solon}}, \bibinfo {author} {\bibfnamefont {Y.}~\bibnamefont {Fily}}, \bibinfo {author} {\bibfnamefont {A.}~\bibnamefont {Baskaran}}, \bibinfo {author} {\bibfnamefont {M.~E.}\ \bibnamefont {Cates}}, \bibinfo {author} {\bibfnamefont {Y.}~\bibnamefont {Kafri}}, \bibinfo {author} {\bibfnamefont {M.}~\bibnamefont {Kardar}},\ and\ \bibinfo {author} {\bibfnamefont {J.}~\bibnamefont {Tailleur}},\ }\bibfield  {title} {\bibinfo {title} {Pressure is not a state function for generic active fluids},\ }\href {https://doi.org/10.1038/nphys3377} {\bibfield  {journal} {\bibinfo  {journal} {Nature physics}\ }\textbf {\bibinfo {volume} {11}},\ \bibinfo {pages} {673} (\bibinfo {year} {2015})}\BibitemShut {NoStop}%
\bibitem [{\citenamefont {Smallenburg}\ and\ \citenamefont {L\"owen}(2015)}]{smallenburg2015swim}%
  \BibitemOpen
  \bibfield  {author} {\bibinfo {author} {\bibfnamefont {F.}~\bibnamefont {Smallenburg}}\ and\ \bibinfo {author} {\bibfnamefont {H.}~\bibnamefont {L\"owen}},\ }\bibfield  {title} {\bibinfo {title} {Swim pressure on walls with curves and corners},\ }\href {https://doi.org/10.1103/PhysRevE.92.032304} {\bibfield  {journal} {\bibinfo  {journal} {Phys. Rev. E}\ }\textbf {\bibinfo {volume} {92}},\ \bibinfo {pages} {032304} (\bibinfo {year} {2015})}\BibitemShut {NoStop}%
\bibitem [{\citenamefont {Ray}\ \emph {et~al.}(2014)\citenamefont {Ray}, \citenamefont {Reichhardt},\ and\ \citenamefont {Reichhardt}}]{ray2014casimir}%
  \BibitemOpen
  \bibfield  {author} {\bibinfo {author} {\bibfnamefont {D.}~\bibnamefont {Ray}}, \bibinfo {author} {\bibfnamefont {C.}~\bibnamefont {Reichhardt}},\ and\ \bibinfo {author} {\bibfnamefont {C.~J.~O.}\ \bibnamefont {Reichhardt}},\ }\bibfield  {title} {\bibinfo {title} {Casimir effect in active matter systems},\ }\href {https://doi.org/10.1103/PhysRevE.90.013019} {\bibfield  {journal} {\bibinfo  {journal} {Phys. Rev. E}\ }\textbf {\bibinfo {volume} {90}},\ \bibinfo {pages} {013019} (\bibinfo {year} {2014})}\BibitemShut {NoStop}%
\bibitem [{\citenamefont {Ni}\ \emph {et~al.}(2015)\citenamefont {Ni}, \citenamefont {Cohen~Stuart},\ and\ \citenamefont {Bolhuis}}]{ni2015tunable}%
  \BibitemOpen
  \bibfield  {author} {\bibinfo {author} {\bibfnamefont {R.}~\bibnamefont {Ni}}, \bibinfo {author} {\bibfnamefont {M.~A.}\ \bibnamefont {Cohen~Stuart}},\ and\ \bibinfo {author} {\bibfnamefont {P.~G.}\ \bibnamefont {Bolhuis}},\ }\bibfield  {title} {\bibinfo {title} {Tunable long range forces mediated by self-propelled colloidal hard spheres},\ }\href {https://doi.org/10.1103/PhysRevLett.114.018302} {\bibfield  {journal} {\bibinfo  {journal} {Phys. Rev. Lett.}\ }\textbf {\bibinfo {volume} {114}},\ \bibinfo {pages} {018302} (\bibinfo {year} {2015})}\BibitemShut {NoStop}%
\bibitem [{\citenamefont {Liu}\ \emph {et~al.}(2020)\citenamefont {Liu}, \citenamefont {Ye}, \citenamefont {Ye}, \citenamefont {Chen},\ and\ \citenamefont {Yang}}]{liu2020constraint}%
  \BibitemOpen
  \bibfield  {author} {\bibinfo {author} {\bibfnamefont {P.}~\bibnamefont {Liu}}, \bibinfo {author} {\bibfnamefont {S.}~\bibnamefont {Ye}}, \bibinfo {author} {\bibfnamefont {F.}~\bibnamefont {Ye}}, \bibinfo {author} {\bibfnamefont {K.}~\bibnamefont {Chen}},\ and\ \bibinfo {author} {\bibfnamefont {M.}~\bibnamefont {Yang}},\ }\bibfield  {title} {\bibinfo {title} {Constraint dependence of active depletion forces on passive particles},\ }\href {https://doi.org/10.1103/PhysRevLett.124.158001} {\bibfield  {journal} {\bibinfo  {journal} {Phys. Rev. Lett.}\ }\textbf {\bibinfo {volume} {124}},\ \bibinfo {pages} {158001} (\bibinfo {year} {2020})}\BibitemShut {NoStop}%
\bibitem [{\citenamefont {Di~Bello}\ \emph {et~al.}(2024)\citenamefont {Di~Bello}, \citenamefont {Majumdar}, \citenamefont {Marathe}, \citenamefont {Metzler},\ and\ \citenamefont {Rold{\'a}n}}]{di2024brownian}%
  \BibitemOpen
  \bibfield  {author} {\bibinfo {author} {\bibfnamefont {C.}~\bibnamefont {Di~Bello}}, \bibinfo {author} {\bibfnamefont {R.}~\bibnamefont {Majumdar}}, \bibinfo {author} {\bibfnamefont {R.}~\bibnamefont {Marathe}}, \bibinfo {author} {\bibfnamefont {R.}~\bibnamefont {Metzler}},\ and\ \bibinfo {author} {\bibfnamefont {{\'E}.}~\bibnamefont {Rold{\'a}n}},\ }\bibfield  {title} {\bibinfo {title} {Brownian particle in a poisson-shot-noise active bath: Exact statistics, effective temperature, and inference},\ }\href {https://doi.org/10.1002/andp.202300427} {\bibfield  {journal} {\bibinfo  {journal} {Annalen der Physik}\ ,\ \bibinfo {pages} {2300427}} (\bibinfo {year} {2024})}\BibitemShut {NoStop}%
\bibitem [{\citenamefont {Rold{\'a}n}\ \emph {et~al.}(2021)\citenamefont {Rold{\'a}n}, \citenamefont {Barral}, \citenamefont {Martin}, \citenamefont {Parrondo},\ and\ \citenamefont {J{\"u}licher}}]{roldan2021quantifying}%
  \BibitemOpen
  \bibfield  {author} {\bibinfo {author} {\bibfnamefont {{\'E}.}~\bibnamefont {Rold{\'a}n}}, \bibinfo {author} {\bibfnamefont {J.}~\bibnamefont {Barral}}, \bibinfo {author} {\bibfnamefont {P.}~\bibnamefont {Martin}}, \bibinfo {author} {\bibfnamefont {J.~M.}\ \bibnamefont {Parrondo}},\ and\ \bibinfo {author} {\bibfnamefont {F.}~\bibnamefont {J{\"u}licher}},\ }\bibfield  {title} {\bibinfo {title} {Quantifying entropy production in active fluctuations of the hair-cell bundle from time irreversibility and uncertainty relations},\ }\href {https://doi.org/10.1088/1367-2630/ac0f18} {\bibfield  {journal} {\bibinfo  {journal} {New Journal of Physics}\ }\textbf {\bibinfo {volume} {23}},\ \bibinfo {pages} {083013} (\bibinfo {year} {2021})}\BibitemShut {NoStop}%
\bibitem [{\citenamefont {Dabelow}\ \emph {et~al.}(2019)\citenamefont {Dabelow}, \citenamefont {Bo},\ and\ \citenamefont {Eichhorn}}]{dabelow2019irreversibility}%
  \BibitemOpen
  \bibfield  {author} {\bibinfo {author} {\bibfnamefont {L.}~\bibnamefont {Dabelow}}, \bibinfo {author} {\bibfnamefont {S.}~\bibnamefont {Bo}},\ and\ \bibinfo {author} {\bibfnamefont {R.}~\bibnamefont {Eichhorn}},\ }\bibfield  {title} {\bibinfo {title} {Irreversibility in active matter systems: Fluctuation theorem and mutual information},\ }\href {https://doi.org/10.1103/PhysRevX.9.021009} {\bibfield  {journal} {\bibinfo  {journal} {Physical Review X}\ }\textbf {\bibinfo {volume} {9}},\ \bibinfo {pages} {021009} (\bibinfo {year} {2019})}\BibitemShut {NoStop}%
\bibitem [{\citenamefont {Manikandan}\ \emph {et~al.}(2021)\citenamefont {Manikandan}, \citenamefont {Ghosh}, \citenamefont {Kundu}, \citenamefont {Das}, \citenamefont {Agrawal}, \citenamefont {Mitra}, \citenamefont {Banerjee},\ and\ \citenamefont {Krishnamurthy}}]{manikandan2021quantitative}%
  \BibitemOpen
  \bibfield  {author} {\bibinfo {author} {\bibfnamefont {S.~K.}\ \bibnamefont {Manikandan}}, \bibinfo {author} {\bibfnamefont {S.}~\bibnamefont {Ghosh}}, \bibinfo {author} {\bibfnamefont {A.}~\bibnamefont {Kundu}}, \bibinfo {author} {\bibfnamefont {B.}~\bibnamefont {Das}}, \bibinfo {author} {\bibfnamefont {V.}~\bibnamefont {Agrawal}}, \bibinfo {author} {\bibfnamefont {D.}~\bibnamefont {Mitra}}, \bibinfo {author} {\bibfnamefont {A.}~\bibnamefont {Banerjee}},\ and\ \bibinfo {author} {\bibfnamefont {S.}~\bibnamefont {Krishnamurthy}},\ }\bibfield  {title} {\bibinfo {title} {Quantitative analysis of non-equilibrium systems from short-time experimental data},\ }\href {https://doi.org/10.1038/s42005-021-00766} {\bibfield  {journal} {\bibinfo  {journal} {Communications Physics}\ }\textbf {\bibinfo {volume} {4}},\ \bibinfo {pages} {258} (\bibinfo {year} {2021})}\BibitemShut {NoStop}%
\bibitem [{\citenamefont {Das}\ \emph {et~al.}(2024)\citenamefont {Das}, \citenamefont {Manikandan}, \citenamefont {Paul}, \citenamefont {Kundu}, \citenamefont {Krishnamurthy},\ and\ \citenamefont {Banerjee}}]{das2024irreversibility}%
  \BibitemOpen
  \bibfield  {author} {\bibinfo {author} {\bibfnamefont {B.}~\bibnamefont {Das}}, \bibinfo {author} {\bibfnamefont {S.~K.}\ \bibnamefont {Manikandan}}, \bibinfo {author} {\bibfnamefont {S.}~\bibnamefont {Paul}}, \bibinfo {author} {\bibfnamefont {A.}~\bibnamefont {Kundu}}, \bibinfo {author} {\bibfnamefont {S.}~\bibnamefont {Krishnamurthy}},\ and\ \bibinfo {author} {\bibfnamefont {A.}~\bibnamefont {Banerjee}},\ }\bibfield  {title} {\bibinfo {title} {Tuning irreversibility of mesoscopic processes using hydrodynamic interactions},\ }\href {https://arxiv.org/abs/2405.00800} {\bibfield  {journal} {\bibinfo  {journal} {arXiv preprint arXiv:2405.00800}\ } (\bibinfo {year} {2024})}\BibitemShut {NoStop}%
\bibitem [{\citenamefont {Sekimoto}(1998)}]{sekimoto1998langevin}%
  \BibitemOpen
  \bibfield  {author} {\bibinfo {author} {\bibfnamefont {K.}~\bibnamefont {Sekimoto}},\ }\bibfield  {title} {\bibinfo {title} {Langevin equation and thermodynamics},\ }\href {https://doi.org/10.1143/PTPS.130.17} {\bibfield  {journal} {\bibinfo  {journal} {Progress of Theoretical Physics Supplement}\ }\textbf {\bibinfo {volume} {130}},\ \bibinfo {pages} {17} (\bibinfo {year} {1998})}\BibitemShut {NoStop}%
\bibitem [{\citenamefont {Das}\ \emph {et~al.}(2023)\citenamefont {Das}, \citenamefont {Paul}, \citenamefont {Manikandan},\ and\ \citenamefont {Banerjee}}]{das2023enhanced}%
  \BibitemOpen
  \bibfield  {author} {\bibinfo {author} {\bibfnamefont {B.}~\bibnamefont {Das}}, \bibinfo {author} {\bibfnamefont {S.}~\bibnamefont {Paul}}, \bibinfo {author} {\bibfnamefont {S.~K.}\ \bibnamefont {Manikandan}},\ and\ \bibinfo {author} {\bibfnamefont {A.}~\bibnamefont {Banerjee}},\ }\bibfield  {title} {\bibinfo {title} {Enhanced directionality of active processes in a viscoelastic bath},\ }\href {J. Phys. 25 (2023) 093051 https://doi.org/10.1088/1367-2630/acfb31} {\bibfield  {journal} {\bibinfo  {journal} {New Journal of Physics}\ }\textbf {\bibinfo {volume} {25}},\ \bibinfo {pages} {093051} (\bibinfo {year} {2023})}\BibitemShut {NoStop}%
\bibitem [{\citenamefont {Ghosh}\ \emph {et~al.}(2025)\citenamefont {Ghosh}, \citenamefont {Majhi}, \citenamefont {Kundu}, \citenamefont {Banerjee},\ and\ \citenamefont {Sinha}}]{Ghosh2025.02.17.638627}%
  \BibitemOpen
  \bibfield  {author} {\bibinfo {author} {\bibfnamefont {T.}~\bibnamefont {Ghosh}}, \bibinfo {author} {\bibfnamefont {A.}~\bibnamefont {Majhi}}, \bibinfo {author} {\bibfnamefont {A.}~\bibnamefont {Kundu}}, \bibinfo {author} {\bibfnamefont {A.}~\bibnamefont {Banerjee}},\ and\ \bibinfo {author} {\bibfnamefont {B.}~\bibnamefont {Sinha}},\ }\bibfield  {title} {\bibinfo {title} {Actin contractility and endocytosis create apico-basal tension gradient in hela cells},\ }\bibfield  {journal} {\bibinfo  {journal} {bioRxiv}\ }\href {https://doi.org/10.1101/2025.02.17.638627} {10.1101/2025.02.17.638627} (\bibinfo {year} {2025})\BibitemShut {NoStop}%
\bibitem [{\citenamefont {Chen}\ \emph {et~al.}(2023)\citenamefont {Chen}, \citenamefont {Yan},\ and\ \citenamefont {Chen}}]{chen2023understanding}%
  \BibitemOpen
  \bibfield  {author} {\bibinfo {author} {\bibfnamefont {J.}~\bibnamefont {Chen}}, \bibinfo {author} {\bibfnamefont {D.}~\bibnamefont {Yan}},\ and\ \bibinfo {author} {\bibfnamefont {Y.}~\bibnamefont {Chen}},\ }\bibfield  {title} {\bibinfo {title} {Understanding the driving force for cell migration plasticity},\ }\href {https://doi.org/10.1016/j.bpj.2023.04.008} {\bibfield  {journal} {\bibinfo  {journal} {Biophysical Journal}\ }\textbf {\bibinfo {volume} {122}},\ \bibinfo {pages} {3570} (\bibinfo {year} {2023})}\BibitemShut {NoStop}%
\bibitem [{\citenamefont {Saraswathibhatla}\ \emph {et~al.}(2020)\citenamefont {Saraswathibhatla}, \citenamefont {Galles},\ and\ \citenamefont {Notbohm}}]{saraswathibhatla2020spatiotemporal}%
  \BibitemOpen
  \bibfield  {author} {\bibinfo {author} {\bibfnamefont {A.}~\bibnamefont {Saraswathibhatla}}, \bibinfo {author} {\bibfnamefont {E.~E.}\ \bibnamefont {Galles}},\ and\ \bibinfo {author} {\bibfnamefont {J.}~\bibnamefont {Notbohm}},\ }\bibfield  {title} {\bibinfo {title} {Spatiotemporal force and motion in collective cell migration},\ }\href {https://doi.org/10.6084/m9.figshare.12378218} {\bibfield  {journal} {\bibinfo  {journal} {Scientific Data}\ }\textbf {\bibinfo {volume} {7}},\ \bibinfo {pages} {197} (\bibinfo {year} {2020})}\BibitemShut {NoStop}%
\bibitem [{\citenamefont {Sabass}\ \emph {et~al.}(2017)\citenamefont {Sabass}, \citenamefont {Koch}, \citenamefont {Liu}, \citenamefont {Stone},\ and\ \citenamefont {Shaevitz}}]{sabass2017force}%
  \BibitemOpen
  \bibfield  {author} {\bibinfo {author} {\bibfnamefont {B.}~\bibnamefont {Sabass}}, \bibinfo {author} {\bibfnamefont {M.~D.}\ \bibnamefont {Koch}}, \bibinfo {author} {\bibfnamefont {G.}~\bibnamefont {Liu}}, \bibinfo {author} {\bibfnamefont {H.~A.}\ \bibnamefont {Stone}},\ and\ \bibinfo {author} {\bibfnamefont {J.~W.}\ \bibnamefont {Shaevitz}},\ }\bibfield  {title} {\bibinfo {title} {Force generation by groups of migrating bacteria},\ }\href {https://doi.org/10.1073/pnas.1621469114} {\bibfield  {journal} {\bibinfo  {journal} {Proceedings of the National Academy of Sciences}\ }\textbf {\bibinfo {volume} {114}},\ \bibinfo {pages} {7266} (\bibinfo {year} {2017})}\BibitemShut {NoStop}%
\bibitem [{\citenamefont {Heisenberg}\ and\ \citenamefont {Bella{\"\i}che}(2013)}]{heisenberg2013forces}%
  \BibitemOpen
  \bibfield  {author} {\bibinfo {author} {\bibfnamefont {C.-P.}\ \bibnamefont {Heisenberg}}\ and\ \bibinfo {author} {\bibfnamefont {Y.}~\bibnamefont {Bella{\"\i}che}},\ }\bibfield  {title} {\bibinfo {title} {Forces in tissue morphogenesis and patterning},\ }\href {https://doi.org/10.1016/j.cell.2013.05.008} {\bibfield  {journal} {\bibinfo  {journal} {Cell}\ }\textbf {\bibinfo {volume} {153}},\ \bibinfo {pages} {948} (\bibinfo {year} {2013})}\BibitemShut {NoStop}%
\bibitem [{\citenamefont {Wiener}(1949)}]{wiener1949extrapolation}%
  \BibitemOpen
  \bibfield  {author} {\bibinfo {author} {\bibfnamefont {N.}~\bibnamefont {Wiener}},\ }\href {https://doi.org/10.7551/mitpress/2946.001.0001} {\emph {\bibinfo {title} {Extrapolation, interpolation, and smoothing of stationary time series: with engineering applications}}}\ (\bibinfo  {publisher} {The MIT press},\ \bibinfo {year} {1949})\BibitemShut {NoStop}%
\bibitem [{\citenamefont {Wiener}(1923)}]{wiener1923differential}%
  \BibitemOpen
  \bibfield  {author} {\bibinfo {author} {\bibfnamefont {N.}~\bibnamefont {Wiener}},\ }\bibfield  {title} {\bibinfo {title} {Differential-space},\ }\href {https://doi.org/10.1002/sapm192321131} {\bibfield  {journal} {\bibinfo  {journal} {Journal of Mathematics and Physics}\ }\textbf {\bibinfo {volume} {2}},\ \bibinfo {pages} {131} (\bibinfo {year} {1923})}\BibitemShut {NoStop}%
\bibitem [{\citenamefont {Boulfelfel}\ \emph {et~al.}(1994)\citenamefont {Boulfelfel}, \citenamefont {Rangayyan}, \citenamefont {Hahn},\ and\ \citenamefont {Kloiber}}]{boulfelfel1994three}%
  \BibitemOpen
  \bibfield  {author} {\bibinfo {author} {\bibfnamefont {D.}~\bibnamefont {Boulfelfel}}, \bibinfo {author} {\bibfnamefont {R.}~\bibnamefont {Rangayyan}}, \bibinfo {author} {\bibfnamefont {L.}~\bibnamefont {Hahn}},\ and\ \bibinfo {author} {\bibfnamefont {R.}~\bibnamefont {Kloiber}},\ }\bibfield  {title} {\bibinfo {title} {Three-dimensional restoration of single photon emission computed tomography images},\ }\href {https://doi.org/10.1109/23.317385} {\bibfield  {journal} {\bibinfo  {journal} {IEEE transactions on nuclear science}\ }\textbf {\bibinfo {volume} {41}},\ \bibinfo {pages} {1746} (\bibinfo {year} {1994})}\BibitemShut {NoStop}%
\bibitem [{\citenamefont {Diniz}(2020)}]{diniz1997adaptive}%
  \BibitemOpen
  \bibfield  {author} {\bibinfo {author} {\bibfnamefont {P.~S.}\ \bibnamefont {Diniz}},\ }\href {https://doi.org/10.1007/978-3-030-29057-3} {\emph {\bibinfo {title} {Adaptive filtering}}},\ Vol.~\bibinfo {volume} {5}\ (\bibinfo  {publisher} {Springer},\ \bibinfo {year} {2020})\BibitemShut {NoStop}%
\bibitem [{\citenamefont {Fricks}\ \emph {et~al.}(2009)\citenamefont {Fricks}, \citenamefont {Yao}, \citenamefont {Elston},\ and\ \citenamefont {Forest}}]{fricks2009time}%
  \BibitemOpen
  \bibfield  {author} {\bibinfo {author} {\bibfnamefont {J.}~\bibnamefont {Fricks}}, \bibinfo {author} {\bibfnamefont {L.}~\bibnamefont {Yao}}, \bibinfo {author} {\bibfnamefont {T.~C.}\ \bibnamefont {Elston}},\ and\ \bibinfo {author} {\bibfnamefont {M.~G.}\ \bibnamefont {Forest}},\ }\bibfield  {title} {\bibinfo {title} {Time-domain methods for diffusive transport in soft matter},\ }\href {https://doi.org/10.1137/070695186} {\bibfield  {journal} {\bibinfo  {journal} {SIAM journal on applied mathematics}\ }\textbf {\bibinfo {volume} {69}},\ \bibinfo {pages} {1277} (\bibinfo {year} {2009})}\BibitemShut {NoStop}%
\bibitem [{\citenamefont {Paul}\ \emph {et~al.}(2021)\citenamefont {Paul}, \citenamefont {Narinder}, \citenamefont {Banerjee}, \citenamefont {Nayak}, \citenamefont {Steindl},\ and\ \citenamefont {Bechinger}}]{paul2021bayesian}%
  \BibitemOpen
  \bibfield  {author} {\bibinfo {author} {\bibfnamefont {S.}~\bibnamefont {Paul}}, \bibinfo {author} {\bibfnamefont {N.}~\bibnamefont {Narinder}}, \bibinfo {author} {\bibfnamefont {A.}~\bibnamefont {Banerjee}}, \bibinfo {author} {\bibfnamefont {K.~R.}\ \bibnamefont {Nayak}}, \bibinfo {author} {\bibfnamefont {J.}~\bibnamefont {Steindl}},\ and\ \bibinfo {author} {\bibfnamefont {C.}~\bibnamefont {Bechinger}},\ }\bibfield  {title} {\bibinfo {title} {Bayesian inference of the viscoelastic properties of a jeffrey’s fluid using optical tweezers},\ }\href {https://doi.org/10.1038/s41598-021-81094-x} {\bibfield  {journal} {\bibinfo  {journal} {Scientific Reports}\ }\textbf {\bibinfo {volume} {11}},\ \bibinfo {pages} {2023} (\bibinfo {year} {2021})}\BibitemShut {NoStop}%
\bibitem [{\citenamefont {MathWorks}()}]{mathworks_varm}%
  \BibitemOpen
  \bibfield  {author} {\bibinfo {author} {\bibnamefont {MathWorks}},\ }\href@noop {} {\bibinfo {title} {Var model (varm)}},\ \bibinfo {howpublished} {\url{https://in.mathworks.com/help/econ/varm.html}}\BibitemShut {NoStop}%
\bibitem [{\citenamefont {Parthasarathy}(2012)}]{parthasarathy2012rapid}%
  \BibitemOpen
  \bibfield  {author} {\bibinfo {author} {\bibfnamefont {R.}~\bibnamefont {Parthasarathy}},\ }\bibfield  {title} {\bibinfo {title} {Rapid, accurate particle tracking by calculation of radial symmetry centers},\ }\href {https://doi.org/10.1038/nmeth.2071} {\bibfield  {journal} {\bibinfo  {journal} {Nature Methods}\ }\textbf {\bibinfo {volume} {9}},\ \bibinfo {pages} {724} (\bibinfo {year} {2012})}\BibitemShut {NoStop}%
\bibitem [{\citenamefont {Das}\ and\ \citenamefont {Manikandan}(2025)}]{das2025localizing}%
  \BibitemOpen
  \bibfield  {author} {\bibinfo {author} {\bibfnamefont {B.}~\bibnamefont {Das}}\ and\ \bibinfo {author} {\bibfnamefont {S.~K.}\ \bibnamefont {Manikandan}},\ }\bibfield  {title} {\bibinfo {title} {Localizing entropy production along non-equilibrium trajectories},\ }\href {https://doi.org/10.48550/arXiv.2503.20427} {\bibfield  {journal} {\bibinfo  {journal} {arXiv preprint arXiv:2503.20427}\ } (\bibinfo {year} {2025})}\BibitemShut {NoStop}%
\bibitem [{\citenamefont {Huang}\ \emph {et~al.}(2025)\citenamefont {Huang}, \citenamefont {Liu}, \citenamefont {Miao},\ and\ \citenamefont {Zhou}}]{huang2025entropy}%
  \BibitemOpen
  \bibfield  {author} {\bibinfo {author} {\bibfnamefont {Y.}~\bibnamefont {Huang}}, \bibinfo {author} {\bibfnamefont {C.}~\bibnamefont {Liu}}, \bibinfo {author} {\bibfnamefont {B.}~\bibnamefont {Miao}},\ and\ \bibinfo {author} {\bibfnamefont {X.}~\bibnamefont {Zhou}},\ }\bibfield  {title} {\bibinfo {title} {Entropy production in non-gaussian active matter: A unified fluctuation theorem and deep learning framework},\ }\href {https://doi.org/10.48550/arXiv.2504.06628} {\bibfield  {journal} {\bibinfo  {journal} {arXiv preprint arXiv:2504.06628}\ } (\bibinfo {year} {2025})}\BibitemShut {NoStop}%
\bibitem [{\citenamefont {Narinder}\ \emph {et~al.}(2019)\citenamefont {Narinder}, \citenamefont {Gomez-Solano},\ and\ \citenamefont {Bechinger}}]{narinder2019active}%
  \BibitemOpen
  \bibfield  {author} {\bibinfo {author} {\bibfnamefont {N.}~\bibnamefont {Narinder}}, \bibinfo {author} {\bibfnamefont {J.~R.}\ \bibnamefont {Gomez-Solano}},\ and\ \bibinfo {author} {\bibfnamefont {C.}~\bibnamefont {Bechinger}},\ }\bibfield  {title} {\bibinfo {title} {Active particles in geometrically confined viscoelastic fluids},\ }\href {https://doi.org/10.1088/1367-2630/ab40e0} {\bibfield  {journal} {\bibinfo  {journal} {New Journal of Physics}\ }\textbf {\bibinfo {volume} {21}},\ \bibinfo {pages} {093058} (\bibinfo {year} {2019})}\BibitemShut {NoStop}%
\bibitem [{\citenamefont {Nakul}\ \emph {et~al.}(2023)\citenamefont {Nakul}, \citenamefont {Roy}, \citenamefont {Nalupurackal}, \citenamefont {Chakraborty}, \citenamefont {Siwach}, \citenamefont {Goswami}, \citenamefont {Edwina}, \citenamefont {Bajpai}, \citenamefont {Singh},\ and\ \citenamefont {Roy}}]{nakul2023studying}%
  \BibitemOpen
  \bibfield  {author} {\bibinfo {author} {\bibfnamefont {U.}~\bibnamefont {Nakul}}, \bibinfo {author} {\bibfnamefont {S.}~\bibnamefont {Roy}}, \bibinfo {author} {\bibfnamefont {G.}~\bibnamefont {Nalupurackal}}, \bibinfo {author} {\bibfnamefont {S.}~\bibnamefont {Chakraborty}}, \bibinfo {author} {\bibfnamefont {P.}~\bibnamefont {Siwach}}, \bibinfo {author} {\bibfnamefont {J.}~\bibnamefont {Goswami}}, \bibinfo {author} {\bibfnamefont {P.}~\bibnamefont {Edwina}}, \bibinfo {author} {\bibfnamefont {S.~K.}\ \bibnamefont {Bajpai}}, \bibinfo {author} {\bibfnamefont {R.}~\bibnamefont {Singh}},\ and\ \bibinfo {author} {\bibfnamefont {B.}~\bibnamefont {Roy}},\ }\bibfield  {title} {\bibinfo {title} {Studying fluctuating trajectories of optically confined passive tracers inside cells provides familiar active forces},\ }\href {https://doi.org/10.1364/BOE.499990} {\bibfield  {journal} {\bibinfo  {journal} {Biomedical optics express}\ }\textbf {\bibinfo {volume} {14}},\ \bibinfo {pages} {5440} (\bibinfo {year}
  {2023})}\BibitemShut {NoStop}%
\bibitem [{\citenamefont {Muenker}\ \emph {et~al.}(2024)\citenamefont {Muenker}, \citenamefont {Knotz}, \citenamefont {Kr{\"u}ger},\ and\ \citenamefont {Betz}}]{muenker2024accessing}%
  \BibitemOpen
  \bibfield  {author} {\bibinfo {author} {\bibfnamefont {T.~M.}\ \bibnamefont {Muenker}}, \bibinfo {author} {\bibfnamefont {G.}~\bibnamefont {Knotz}}, \bibinfo {author} {\bibfnamefont {M.}~\bibnamefont {Kr{\"u}ger}},\ and\ \bibinfo {author} {\bibfnamefont {T.}~\bibnamefont {Betz}},\ }\bibfield  {title} {\bibinfo {title} {Accessing activity and viscoelastic properties of artificial and living systems from passive measurement},\ }\href {https://doi.org/10.1038/s41563-024-01957-2} {\bibfield  {journal} {\bibinfo  {journal} {Nature Materials}\ }\textbf {\bibinfo {volume} {23}},\ \bibinfo {pages} {1283} (\bibinfo {year} {2024})}\BibitemShut {NoStop}%
\bibitem [{\citenamefont {Pahi}\ \emph {et~al.}(2025)\citenamefont {Pahi}, \citenamefont {Sahoo}, \citenamefont {Das}, \citenamefont {Paul},\ and\ \citenamefont {Banerjee}}]{pahi2025simultaneous}%
  \BibitemOpen
  \bibfield  {author} {\bibinfo {author} {\bibfnamefont {A.}~\bibnamefont {Pahi}}, \bibinfo {author} {\bibfnamefont {K.~R.}\ \bibnamefont {Sahoo}}, \bibinfo {author} {\bibfnamefont {B.}~\bibnamefont {Das}}, \bibinfo {author} {\bibfnamefont {S.}~\bibnamefont {Paul}},\ and\ \bibinfo {author} {\bibfnamefont {A.}~\bibnamefont {Banerjee}},\ }\bibfield  {title} {\bibinfo {title} {Simultaneous active and diffusive behaviour of asymmetric microclusters in a photophoretic trap},\ }\href {https://doi.org/10.48550/arXiv.2503.16059} {\bibfield  {journal} {\bibinfo  {journal} {arXiv preprint arXiv:2503.16059}\ } (\bibinfo {year} {2025})}\BibitemShut {NoStop}%
\bibitem [{\citenamefont {Gupta}\ \emph {et~al.}(2022)\citenamefont {Gupta}, \citenamefont {Khan}, \citenamefont {Biswas}, \citenamefont {Mondal}, \citenamefont {Das}, \citenamefont {Sharif},\ and\ \citenamefont {Mallick}}]{gupta2022combined}%
  \BibitemOpen
  \bibfield  {author} {\bibinfo {author} {\bibfnamefont {S.}~\bibnamefont {Gupta}}, \bibinfo {author} {\bibfnamefont {A.}~\bibnamefont {Khan}}, \bibinfo {author} {\bibfnamefont {P.}~\bibnamefont {Biswas}}, \bibinfo {author} {\bibfnamefont {K.}~\bibnamefont {Mondal}}, \bibinfo {author} {\bibfnamefont {D.}~\bibnamefont {Das}}, \bibinfo {author} {\bibfnamefont {S.}~\bibnamefont {Sharif}},\ and\ \bibinfo {author} {\bibfnamefont {A.~I.}\ \bibnamefont {Mallick}},\ }\bibfield  {title} {\bibinfo {title} {A combined protocol for isolation of t6ss-positive campylobacter jejuni and assessment of interspecies interaction},\ }\href {https://doi.org/10.1016/j.xpro.2022.101368} {\bibfield  {journal} {\bibinfo  {journal} {STAR protocols}\ }\textbf {\bibinfo {volume} {3}},\ \bibinfo {pages} {101368} (\bibinfo {year} {2022})}\BibitemShut {NoStop}%
\bibitem [{\citenamefont {Seyforth}\ \emph {et~al.}(2022)\citenamefont {Seyforth}, \citenamefont {Gomez}, \citenamefont {Rogers}, \citenamefont {Ross},\ and\ \citenamefont {Ahmed}}]{seyforth2022nonequilibrium}%
  \BibitemOpen
  \bibfield  {author} {\bibinfo {author} {\bibfnamefont {H.}~\bibnamefont {Seyforth}}, \bibinfo {author} {\bibfnamefont {M.}~\bibnamefont {Gomez}}, \bibinfo {author} {\bibfnamefont {W.~B.}\ \bibnamefont {Rogers}}, \bibinfo {author} {\bibfnamefont {J.~L.}\ \bibnamefont {Ross}},\ and\ \bibinfo {author} {\bibfnamefont {W.~W.}\ \bibnamefont {Ahmed}},\ }\bibfield  {title} {\bibinfo {title} {Nonequilibrium fluctuations and nonlinear response of an active bath},\ }\href {https://doi.org/10.1103/PhysRevResearch.4.023043} {\bibfield  {journal} {\bibinfo  {journal} {Physical Review Research}\ }\textbf {\bibinfo {volume} {4}},\ \bibinfo {pages} {023043} (\bibinfo {year} {2022})}\BibitemShut {NoStop}%
\end{thebibliography}

%

\nocite{}
\end{document}